\input harvmac.tex
\input amssym.def
\input amssym.tex

\def\unlockat{\catcode`\@=11}
\def\lockat{\catcode`\@=12}
\unlockat
\def\newsec#1{\global\advance\secno by1\message{(\the\secno. #1)}
\global\subsecno=0\global\subsubsecno=0
\global\deno=0\global\prono=0\global\teno=0\eqnres@t\noindent
{\bf\the\secno. #1} \writetoca{{\secsym}
{#1}}\par\nobreak\medskip\nobreak}
\global\newcount\subsecno \global\subsecno=0
\def\subsec#1{\global\advance\subsecno by1\message{(\secsym\the\subsecno.
#1)}
\ifnum\lastpenalty>9000\else\bigbreak\fi\global\subsubsecno=0
\global\deno=0\global\prono=0\global\teno=0
\noindent{\it\secsym\the\subsecno. #1} \writetoca{\string\quad
{\secsym\the\subsecno.} {#1}}
\par\nobreak\medskip\nobreak}
\global\newcount\subsubsecno \global\subsubsecno=0
\def\subsubsec#1{\global\advance\subsubsecno by1
\message{(\secsym\the\subsecno.\the\subsubsecno. #1)}
\ifnum\lastpenalty>9000\else\bigbreak\fi
\noindent\quad{\secsym\the\subsecno.\the\subsubsecno.}{#1}
\writetoca{\string\qquad{\secsym\the\subsecno.\the\subsubsecno.}{#1}}
\par\nobreak\medskip\nobreak}
\global\newcount\deno \global\deno=0
\def\de#1{\global\advance\deno by1
\message{(\bf Definition\quad\secsym\the\subsecno.\the\deno #1)}
\ifnum\lastpenalty>9000\else\bigbreak\fi
\noindent{\bf Definition\quad\secsym\the\subsecno.\the\deno}{#1}
\writetoca{\string\qquad{\secsym\the\subsecno.\the\deno}{#1}}}
\global\newcount\prono \global\prono=0
\def\pro#1{\global\advance\prono by1
\message{(\bf Proposition\quad\secsym\the\subsecno.\the\prono #1)}
\ifnum\lastpenalty>9000\else\bigbreak\fi
\noindent{\bf Proposition\quad\secsym\the\subsecno.\the\prono}{#1}
\writetoca{\string\qquad{\secsym\the\subsecno.\the\prono}{#1}}}
\global\newcount\teno \global\prono=0
\def\te#1{\global\advance\teno by1
\message{(\bf Theorem\quad\secsym\the\subsecno.\the\teno #1)}
\ifnum\lastpenalty>9000\else\bigbreak\fi
\noindent{\bf Theorem\quad\secsym\the\subsecno.\the\teno}{#1}
\writetoca{\string\qquad{\secsym\the\subsecno.\the\teno}{#1}}}
\def\subsubseclab#1{\DefWarn#1\xdef
#1{\noexpand\hyperref{}{subsubsection}%
{\secsym\the\subsecno.\the\subsubsecno}%
{\secsym\the\subsecno.\the\subsubsecno}}%
\writedef{#1\leftbracket#1}\wrlabeL{#1=#1}}
\lockat

\def\IC{{\Bbb C}}

\def\IQ{{\Bbb Q}}
\def\IR{{\Bbb R}}
\def\IZ{{\Bbb Z}}

\def\pr {\partial}
\def\apr {\overline {\partial }}

\def\refb[1] {{(\ref{\#1})}}
\def\eq[1] {{eq.(\ref{\#1}) }}
\def\bear {{\begin{array}}}

\def\qq[1] {{\frac{\overline{\pr ^2 {\cal F}}}{\pr z^{\overline {\#1}}}}}

\def\inv[1] {{{\#1}^{-1}}} 

\def\LL {{\cal L}}

\def\pr {{\partial}}
\def\apr {{\bar \partial}}

\def\a {{\alpha}}
\def\b {{\beta}}

\def\s {{\sigma}}
\def\l {{\lambda}}

\def\CA {{\cal A}}

\def\CD {{\cal D}}
\def\CE {{\cal E}}
\def\CF {{\cal F}}
\def\CG {{\cal G}}
\def\CH {{\cal H}}

\def\CL {{\cal L}}
\def\CM {{\cal M}}

\def\CO {{\cal O}}

\def\CR {{\cal R}}


\def\Fg{{\frak g}}
\def\Fh{{\frak h}}



\def\zb {\bar{z}}

\def\Tr{{\rm Tr}}

\def\Vol{{\rm Vol}}

\def\ch{{\rm ch}}

\def\Lie{{\rm Lie}}

\font\manual=manfnt \def\dbend{\lower3.5pt\hbox{\manual\char127}}

\def\boxit#1{\vbox{\hrule\hbox{\vrule\kern8pt
\vbox{\hbox{\kern8pt}\hbox{\vbox{#1}}\hbox{\kern8pt}}
\kern8pt\vrule}\hrule}}
\def\mathboxit#1{\vbox{\hrule\hbox{\vrule\kern8pt\vbox{\kern8pt
\hbox{$\displaystyle #1$}\kern8pt}\kern8pt\vrule}\hrule}}
\def\frac#1#2{{#1\over#2}}

\lref\mns{G.~Moore, N.~Nekrasov and S.~Shatashvili, {\it Integrating over
Higgs
Branches, }
Commun.\ Math.\ Phys.\ {\bf 209} (2000) 97,
[arXiv:hep-th/9712241].
}
\lref\GKLO{
  A.~Gerasimov, S.~Kharchev, D.~Lebedev and S.~Oblezin,
   {\it On a class of representations of the Yangian and moduli space of
  monopoles},
  Commun.\ Math.\ Phys.\  {\bf 260}, 511 (2005), [arXiv:
math.AG/0409031].}
\lref\GKLOone{ A.~Gerasimov, S.~Kharchev, D.~Lebedev and S.~Oblezin,
{\it On a Class of Representations of Quantum Groups},
[arXiv:math/0501473].}
\Title{ \vbox{\baselineskip12pt \hbox{hep-th/0609024}
\hbox{HMI-06-03}
\hbox{ITEP-TH-06-43}
\hbox{TCD-MATH-06-11}}}
 {\vbox{
\centerline{Higgs Bundles, Gauge Theories }
\vskip 0.5cm
\centerline{and} \vskip 0.5cm \centerline{Quantum Groups}
\medskip
 }}
\medskip
\centerline{\bf Anton A. Gerasimov $^{1,2,3}$ and Samson L.
Shatashvili $^{2,3,4}$}
\vskip 0.3cm
\centerline{\it $^{1}$ Institute for Theoretical and
Experimental Physics, Moscow, 117259, Russia} \centerline{\it
$^{2}$ School of Mathematics, Trinity
College, Dublin 2, Ireland } \centerline{\it $^3$ The Hamilton
Mathematics Institute TCD, Dublin 2, Ireland}
\centerline{\it $^{4}$ IHES, 35 route de Chartres,
Bures-sur-Yvette, FRANCE}
\vskip 0.3cm
 The appearance of the Bethe Ansatz equation
for the Nonlinear
Schr\"{o}dinger equation in the equivariant integration
over the moduli space of Higgs bundles is  revisited.
We argue  that the wave functions of the
corresponding  two-dimensional  topological $U(N)$ gauge theory
 reproduce  quantum wave functions of the Nonlinear
Schr\"{o}dinger equation  in the $N$-particle sector.
This implies the full equivalence between the above gauge theory
and the $N$-particle sub-sector of the quantum theory of Nonlinear
 Schr\"{o}dinger equation. This also implies the explicit
 correspondence
between the gauge theory and the representation theory
of degenerate double affine Hecke algebra.
We  propose similar construction based on the
$G/G$ gauged WZW
model leading to
the representation theory of the
double affine Hecke algebra.
 The relation with the Nahm transform and
the geometric Langlands correspondence is briefly discussed.
\Date{}
\newsec{Introduction}

In \mns\  a  relation between a certain type of two-dimensional
Yang-Mills theory and the Bethe Ansatz equations for the quantum theory of
the Nonlinear
Schr\"{o}inger  equation was uncovered. The topological
Yang-Mills-Higgs theory considered in \mns\
captures the hyperk\"{a}hler geometry of the moduli space of
Higgs  bundles introduced in \ref\Hitchin{
N. J.~Hitchin, {\it The self-duality equations on a Riemann surface},
Proc. London Math. Soc. 55 (1987), 59-126.} by Hitchin.
It was shown that the path integral in  this theory  can be localized
on  the disconnected set whose components are  naturally
enumerated by the solutions of a system of the Bethe Ansatz  equations.
The conceptual explanation of the appearance of the Bethe Ansatz equations
 as a result of the  localization  in the topological Yang-Mills-Higgs
 theory, as well as potential consequences were missing.

To elucidate the structure of the theory we consider  the space of
 wave functions of the  two-dimensional    gauge theory
 introduced in \mns. We  argue that  this space
can be identified with the space of  wave-functions in the $N$-particle
sector of the quantum theory of the Nonlinear Schr\"{o}dinger
equation constructed in the framework of the coordinate Bethe Ansatz (see
e.g.
 \ref\Ga{M.~Gaudin, {\it La Fonction d'Onde de Bethe},  Paris,
  Masson, 1983.},
\ref\IKS{N.~Bogoliubov, A.~Izergin and V.~Korepin, {\it
Quantum Inverse Scattering Method and Correlation Functions},
Cambridge University Press, Cambridge, England, 1993.}).
This implies the equivalence between two seemingly different
quantum field theories.  Taking into account the interpretation
of the coordinate Bethe Ansatz in the Nonlinear Schr\"{o}dinger
theory via the Quantum Inverse Scattering Method (algebraic Bethe Ansatz)
\ref\SFT{L. D. Faddeev, E. K. Sklyanin and L. A. Takhtajan,
{\it  The Quantum Inverse Problem Method. 1.,}
Theor. Math. Phys. 40: 688-706, 1980, Teor. Mat. Fiz. 40: 194-220, 1979.}
 this also  provides a direct correspondence between
gauge theory based on the moduli problem for Higgs bundles and
the representation theory of  quantum groups.

The partition function  of the theory considered in \mns\
depends on an additional parameter playing the role of the first
Chern class of the tautological line bundle on the classifying
space $BU(1)$ of the $U(1)$ group in the description of the
equivariant cohomology of the Hitchin moduli space. In the the Nonlinear
Schr\"{o}dinger  theory  the same parameter plays the role of the coupling
constant.

In addition we show that the considerations of \mns\ can be
generalized from the Yang-Mills-Higgs theory to  $G/G$ gauged WZW
model. The corresponding partition function is expressed in terms of
  solutions of  Bethe Ansatz equations  similar to the ones
for XXZ  spin chains. Thus presumably the  wave functions
for $G=U(N)$ can be identified with the wave functions of  the
spin chains.  One can suspect that the relations discussed in
this paper are more general  and  other examples considered
in \mns\ have similar interpretation in
terms of the representation theory of quantum groups
\foot{ See in this respect \GKLO,\GKLOone\ where the role of the moduli
spaces
of monopoles in the representation theory of
infinite-dimensional quantum groups was studied.}.
 The case of  the instanton moduli space in the
 four-dimensional  Yang-Mills  theory studied in \mns\
looks especially  interesting in this regard and  will be considered
elsewhere.

Let us finally note
 that the revealed correspondence between topological quantum field
theories and integrable structures captured by
the Bethe Ansatz equations obviously imply some relation with the
old standing wish to unify three-dimensional  hyperbolic
geometry, two dimensional conformal/integrable theories and
(some fragments of ) algebraic K-theory
 \ref\ZAG{
W. D.~Neumann, D.~Zagier, {\it Volumes of hyperbolic 3-manifolds},
 Topology {\bf 24} (1985), 307-332.},
\ref\Nhamone{W.~Nahm,
{\it Conformal field theory, dilogarithms, and three dimensional
  manifolds},
 in:Interface between physics and mathematics,
Proceedings, Hangzhou 1993, W. Nahm and J.M. Shen eds., World
Scientific.},
\ref\GLTATEO{F.~Gliozzi, R.~Tateo, {\it
Thermodynamic Bethe Ansatz and Threefold Triangulations},
 Int. J. of Mod. Phys A, {\bf 11} (1996) 4051,
 [arXiv:hep-th/9505102].},
\ref\Nahm{W.~Nahm,
{\it Conformal Field Theory and Torsion Elements of the Bloch group
}, [arXiv:hep-th/0404120].}.

The plan of the paper is as follows.  In Section 2 we recall
the standard facts from the two-dimensional  Yang-Mills theory and the
$G/G$ gauged WZW  model. This provides a template for further
considerations in the Yang-Mills-Higgs theory. This part can be skipped
 by the reader familiar with the subject.
 In Section 3, following \mns, the  topological Yang-Mills-Higgs theory is
 introduced and the application of  the cohomological localization
 technique is discussed.
We also provide the explicit description of the two important limiting
cases: $c\to \infty$ and $c\to 0$ - these limiting cases are instructive
 in establishing the correspondence between answers computed in this model
and  representation theory.
 In Section 4 we recall the description of exact $N$-particle wave
functions in the quantum theory of the Nonlinear Schr\"{o}dinger equation
emphasizing
the role of the degenerate double affine Hecke algebra. We also consider
the
same limiting cases: $c\to \infty$ and $c\to 0$.
  In addition, we stress the fact  that
the quantum wave functions of Nonlinear Schr\"{o}dinger equation are
$p \rightarrow 1$ limits of the (generalized)
 spherical functions for $GL(N,\IQ_p)$ with $p$ - prime
given by Hall-Littlewood polynomials.
In Section 5  we propose the explicit expressions for the exact wave
functions in the
Yang-Mills-Higgs theory and identify the bases of wave functions with
 the bases of eigenfunctions of the Hamiltonian in the
finite-particle sector for Nonlinear Schr\"{o}dinger equation.
This provides  the conceptual explanation of the appearance
of the Bethe Ansatz equations in \mns.
In Section 6 we give the equivariant cohomology description of the
Hilbert space in the topological Yang-Mills-Higgs theory.
In Sections 7 we discuss the connection between the gauge theory and the
quantum the Nonlinear  Schr\"{o}dinger integrable system
using the Nahm duality and comment on the relation
with the geometric Langlands correspondence.
In Section 8 the relevant generalizations of $G/G$ gauge
Wess-Zumino-Witten  is proposed and partition function
is derived.  We conclude with the discussion of the
possible general framework.

We include two  relevant topics in the Appendices. In  Appendix A
we provide  the twistor type description for the Yang-Mills-Higgs theory
and Appendix B is devoted to the topic of quantization
of a singular manifold relevant to the construction of
the wave function in the main part of the paper.

\newsec{Two-dimensional  gauge theories with compact group}

There is an interesting class of two-dimensional gauge theories that
are exactly solvable on an arbitrary Riemann surface.
 The simplest examples are  given by the  Yang-Mills theory
on a Riemann surface $\Sigma$ with a gauge group
$\CG_{\Sigma}=Map(\Sigma,G)$
with $G$ - a compact Lie  group
\ref\mda{A.~Migdal, {\it Recursion Equations in Gauge Theories},
 Zh. Eksp. Teor. Fiz {\bf 69} (1975) 810
 (Sov. Phys. Jetp. {\bf 42} (1975) 413).
}, \ref\kaz{V. Kazakov and I. Kostov, {\it Nonlinear Strings In
Two-Dimensional U(Infinity) Gauge Theory,} Nucl. Phys. B176 (1980)
199.},\ref\kazk{V. Kazakov, {\it Wilson Loop Average For An Arbitrary
Contour In
Two-Dimensional,}, Nucl.Phys. B {\bf 179} (1981) 283.}, \ref\rus{ B.
Rusakov, {\it Loop Averages And Partition Functions In U(N) Gauge
Theory On Two-Dimensional Manifolds,} Mod.Phys.Lett. A5 (1990)
693.},  \ref\Wone{
  E.~Witten,
  {\it On quantum gauge theories in two-dimensions},
  Commun.\ Math.\ Phys.\  {\bf 141}, 153 (1991).}, \ref\Wtwo{
  E.~Witten,
  {\it Two-dimensional gauge theories revisited},
  J.\ Geom.\ Phys.\  {\bf 9}, 303 (1992)
  [arXiv:hep-th/9204083].},
 and more generally by the $G/G$ gauged WZW theory
\ref\Gawedzki{
  K.~Gawedzki and A.~Kupiainen,
  {\it $G/H$ Conformal Field Theory form Gauged WZW Model},
  Phys.\ Lett.\ B {\bf 215}, 119 (1988).}, \ref\GWZW{
  M.~Spiegelglas and S.~Yankielowicz,
  {\it $G / G$ Topological Field Theories By Cosetting G(K)},
  Nucl.\ Phys.\ B {\bf 393}, 301 (1993)
  [arXiv:hep-th/9201036].}.
 The Lagrangian of the $G/G$ gauged WZW  model
 depends on  an integer number $k$ and the two-dimensional
Yang-Mills theory can be recovered in the limit $k\rightarrow \infty$. In
all
  these cases the space of the classical  solutions of the theory is
naturally described in terms of the moduli spaces associated with the
underlying  Riemann surface and
the  partition function can be represented as a sum over the
 set of irreducible representations of some algebraic object. For the
Yang-Mills theory it is the set of finite-dimensional representations
 of the group $G$ and for the $G/G$ gauged WZW  model it is the set of
finite-dimensional irreducible representations of the quantum
group $U_q{\frak g}$ with
${\frak g}=\Lie(G)$ and $q=exp(2\pi i/(k+c_v))$, $c_v$ - the dual Coxeter
number. In this section we recall the standard facts about
these two-dimensional gauge theories.

\subsec{Two-dimensional  Yang-Mills theory}

 We start  with a two-dimensional  Yang-Mills theory for a compact group
$G$  on a Riemann surface $\Sigma$ (we mostly follow \Wone).
 The partition function is given by:
  \eqn\FInt{
Z_{YM}(\Sigma)=\frac{1}{\Vol(\CG_{\Sigma})}\int \, D\varphi\,DA\,D\psi\,\,
e^{\frac{1}{2\pi}\int_{\Sigma}\,
d^2z\,(i\Tr\,\varphi\,F(A)+\frac{1}{2}\Tr\,\psi \wedge \psi
-g_{YM}^2\Tr\,\varphi^2\,{\rm vol}_{\Sigma})}\,\,\, ,}
where $\Vol(\CG_{\Sigma})$ is a volume of the gauge group
$\CG_{\Sigma}={\rm Map}(\Sigma,G)$ and ${\rm vol}_{\Sigma}$ is a volume
form on $\Sigma$.
Here $A$ is a connection on a principal $G$-bundle over
$\Sigma$, $\varphi\in \CA^0(\Sigma,{\rm ad}_\Fg)$ is a
section   of the  vector bundle ${\rm ad}_\Fg$   and $\psi\in
\CA^1(\Sigma,{\rm ad}_{\Fg})$ is an odd one-form taking values in
${\rm ad}_\Fg$, $\Fg={\rm Lie}(G)$. The measure $DA\,D\psi$ is
a canonical flat measure  and the measure $D\varphi$ is defined using the
standard normalization of the Killing form on $\Fg$.
We also imply the sum over all topological classes of
the principal $G$-bundle over $\Sigma_h$.

The Feynman path integral \FInt\ is invariant under the action of the
following  odd
and  even vector fields:
\eqn\BRSTYM{
Q\,A=i\psi,\,\,\,\,\,\,Q\,\psi=-(d\varphi+[A,\varphi]),\,\,\,\,\,\,\,\,Q\,\varphi=0,}
\eqn\GAUGEYM{
\CL_{\varphi}\,A=(d\varphi+[A,\varphi]),\,\,\,\,\,\CL_{\varphi}\,\psi=-[\varphi,\psi]
,\,\,\,\,\CL_{\varphi}\,\varphi=0,}
such that $Q^2=-i\CL_{\varphi}$. The invariance of the path integral under
 transformations  \BRSTYM, \GAUGEYM\ allows to solve the theory
exactly. We will be  mostly interested in the (generalizations of) the
two-dimensional Yang-Mills theory with $g_{YM}^2=0$.
 The theory for $g_{YM}^2=0$ will be called topological
Yang-Mills theory because its correlation functions
depend only on the topology of the underlying surface.

The observables in the theory are constructed using the standard
descent procedure. Given any ${\rm Ad}_G$-invariant function $f\in
Fun(\Fg)$  the corresponding local observable is  given by:
\eqn\OBSLOC{
\CO^{(0)}_f={\rm Tr}\,f(\varphi),}
where trace is taken in the adjoint representation.
The corresponding nonlocal observables are given by the solutions of
the descent equations:
\eqn\DESCENT{
d\CO^{(i)}=-iQ(\CO^{(i+1)}).}
Thus we have:
\eqn\OBONE{
\CO^{(1)}_f=\int_{\Gamma}
 dz \sum_{a=1}^{{\rm rank}(\Fg)}\,\frac{\pr f(\varphi)}{\pr \varphi^a}
\psi^a,}
\eqn\OBTWO{
\CO^{(2)}_f=\frac{1}{2}
\int_{\Sigma} d^2z \sum_{a,b=1}^{{\rm rank}(\Fg)}\,\frac{\pr^2
f(\varphi)}{\pr
  \varphi^a \,\pr \varphi^b}\psi^a\wedge\psi^b+i
\int_{\Sigma} d^2z \sum_{a=1}^{{\rm rank}(\Fg)}\,\frac{\pr f(\varphi)}{\pr
\varphi^a}
F(A)^a,}
where $\varphi=\sum_{a=1}^{{\rm rank}(\Fg)}\varphi^a\,t_a$, $\{t^a\}$
is a bases in $\Fg$ and $\Gamma\in \Sigma$ is a closed curve on a
two-dimensional
surface $\Sigma$.

To describe the Hilbert space of the the Yang-Mills theory consider
the theory on a two dimensional torus  $\Sigma=T^2$
supplied with the flat coordinates $(t,\s)\sim (t+2\pi m,\sigma+2\pi n)$,
$n,m\in \IZ$. The action has the form:
\eqn\actone{ S(\varphi,A)=\frac{1}{2\pi}\int_{T^2}\, dt d\s\,\,
\Tr\,(\varphi
\pr_t A_{\s}+ A_t(\pr_{\s}\varphi +[A_{\s}, \varphi])).}
Here we have integrate out fermionic degrees of freedom using the
appropriate
choice of the measure in the path integral. Due to the gauge
invariance generated by the first-class constraint:
\eqn\constr{
\pr_{\s}\varphi+[A_{\s},\varphi]=0,}
the phase space of the theory can be reduced to the finite-dimensional
space - the phase space is an orbifold:
 \eqn\PSORB{ \CM_G=(T^*H)/W.} Its non-singular open part
is given by:
 \eqn\PSopen{ \CM_G^{(0)}=(T^*H_0)/W,} where $H_0=H\cap
G^{reg}$ is an intersection of the Cartan torus $H\subset G$ with a
subset $G^{reg}$ of regular elements of $G$ (the set of the elements
of $G$ such that its centralizer has the dimension equal to the rank
of $\Fg$) and $W$ - is Weyl group.
Note that Cartan torus is given by $H=\Fh/Q^{\vee}$ where $Q^{\vee}$
is a coroot lattice of  $\Fg$. In the case of $G=U(N)$ the set
$H/H_0=\cup_{j<k}\{e^{2\pi i x_j}=e^{2\pi ix_k}\}$ is the   main diagonal.
Here $x=\sum_{j=1}^{{\rm rank}(\Fg)}x_je^j$ and $\{e^j\}$ is an
orthonormal bases of $\Fh$.

The Hilbert space of the theory can be realized as a space of  ${\rm
Ad}_G$-invariant  functions on $G$. The paring is defined by the
integration with a
 bi-invariant  normalized Haar  measure:
\eqn\PAIR{
  <\Psi_1,\Psi_2>=\int_G dg\, \overline{\Psi}_1(g)\,\Psi_2(g).}
Being restricted to the subspace of $Ad_G$-invariant functions
it descends to the integral over Cartan torus $H$:
\eqn\REDPAIR{
<\Psi_1,\Psi_2>=\frac{1}{|W|}\int_H dx\,\Delta_G^2(e^{2\pi i x})
\overline{\Psi}_1(x)\,\Psi_2(x),}
where the Jacobian is given by:
 \eqn\Jacob{
\Delta_G^2(e^{2\pi i x})=\prod_{\a \in
R^+}(e^{i\pi\a(x)}-e^{-i\pi\a(x)})^2,}
and $R^+$ is a set of  positive roots of $\Fg$.

The  set of invariant operators descending onto $\CM$ includes a
commuting family of the operators given by
${\rm Ad}_G$-invariant polynomials of $\varphi$. In the case of
$G=U(N)$ one can take:
\eqn\casimirs{
\CO^{(0)}_k(\varphi)=\frac{1}{(2\pi  )^k}\Tr\,\varphi^k.}
where trace is taken in the $N$-dimensional representation.
Define the  bases of wave-functions by  the condition:
\eqn\EIGEN{
\CO^{(0)}_k(\varphi)\Psi_{\l}(x_1,\cdots,x_N)=p_k(\l)\Psi_{\l}(x_1,\cdots,x_N),}
$$
\Psi_{\l}(x_{w(1)},\cdots,x_{w(N)})=
\Psi_{\l}(x_1,\cdots,x_N),\,\,\,\,\,\,w\in W.$$
 where $\l=(\l_1,\cdots,\l_N)$ are elements of the  weight lattice $P$ of
$\Fg$ and $p_k\in \IC[\Fh^*]^W$ is the bases of invariant polynomials
 on the dual $\Fh^*$ to Cartan subalgebra $\Fh$.

It is useful  to redefine  wave functions by multiplying
them on: \eqn\Weylnomod{
\Delta_G(e^{2\pi i x})=\prod_{\a \in R^+}
(e^{i\pi\a(x)}-e^{-i\pi\a(x)}),}
so that the integration measure becomes a flat measure on $H$:
\eqn\FALTMES{<\Psi_1,\Psi_2>=\frac{1}{|W|}\int_H dx\,
\overline{\Phi}_1(x)\,\Phi_2(x),}
where $\Phi_i(x)=\Delta_G(e^{2\pi i x})\Psi_i(x)$.
Then  the action of the  operators $c_k$ on the wave-functions
has a simple form:  \eqn\REDCAS{
\CO^{(0)}_k=\frac{1}{(2\pi i )^k}\sum_{i=1}^N\frac{\pr^k}{\pr x_i^k}.}
Note that after the redefinition the wave functions are skew-symmetric
 with respect to the action of $W$.  To get the symmetric
  wave functions one can use  slightly different
 redefinition $\Phi_i(x)=|\Delta_G(e^{2\pi i x})|\Psi_i(x)$.

The space of states of the theory allows simple description
in terms of the representation theory of $G$.
The bases of  $W$-skew-invariant eigenfunctions
 can be expressed through the characters
$\ch_{\mu}(g)=\Tr_{V_{\mu}} \,g$ of the finite-dimensional
 irreducible representations of $G$  as:
\eqn\RENWAVE{
\Phi_{\mu+\rho}(g)=\Delta_G(e^{2\pi i
x})\,\ch_{\mu}(g),\,\,\,\,\,\,\,\,\,\,\,\,\mu\in
P_{++},}
where $P_{++}$ is a subset of the dominant weights of $G$,
$\rho=\frac{1}{2}\sum_{\a\in R^+} \a$ is a half-sum of  positive
roots and $\l=\mu+\rho$ in \EIGEN.
Explicitly we have:  \eqn\wavefunc{
\Phi_{\mu+\rho}(x)=\sum_{w\in W}(-1)^{l(w)}e^{2\pi i \,w(\mu+\rho)(x)},}
where  $l(w)$ is a length of a reduced decomposition of $w\in W$.

The partition function of the topological
Yang-Mills theory  on a Riemann surface $\Sigma_h$ of a genus
$h$ can be expressed as a sum over unitary irreducible representations of
$G$:
\eqn\PARTFUNC{
Z_{YM}(\Sigma_h)=\left(\frac{{\rm Vol}(G)}{(2\pi)^{{\rm
dim}(G)}}\right)^{2h-2}
\sum_{\mu\in P_{++}} (\dim\, V_{\mu})^{2-2h},}
where: \eqn\DIM{
\dim\, V_{\mu}=\prod_{\a\in R_+}\frac{(\mu+\rho,\a)}{(\rho,\a)},}
is a dimension of the irreducible representation $V_{\mu}$ given by the
Weyl formula. Here $(\a,\b)$ is an invariant symmetric pairing on
$\Fh^*$. Thus for $\Fg={\frak  u}_N$ we have: \eqn\DIMGLN{
\dim\, V_{\mu}=\prod_{1\leq  i<j\leq N}\frac{(\mu_i-\mu_j+j-i)}{(j-i)},}
where  $\mu_i\in \IZ_+$ and  $\mu_1\geq \mu_2\geq \cdots \geq \mu_N$.

Note that for $h=0,1$  the partition function \PARTFUNC\ is divergent.
It can be made finite by taking   $g_{YM}\neq 0$ in \FInt. We use
more general regularization compatible with symmetries of the theory -
we consider the following deformation of the action:
\eqn\DEFACTN{
\Delta S_{YM}=-\sum_{k=1}^{\infty}
\,t_k\,\,\int_{\Sigma_h}d^2z \,\,\CO^{(0)}_k(\varphi)\,{\rm
vol}_{\Sigma_h}}
where $\CO^{(0)}_k(\varphi)$ is a bases of ${\rm Ad}_{\Fg}$-invariant
polynomials
on $\varphi$ and the number of $t_k\neq 0$ is finite.
We also impose the additional condition on $t_k$ such that the
functional integral is well defined.   For  $\Fg={\frak u}_N$ one can
take $\CO^{(0)}_k(\varphi)=\frac{1}{(2\pi )^k}\Tr\,\varphi^{k}$.
Then the partition function is given by:
\eqn\REGULPART{
Z_{YM}(\Sigma_h)=\left(\frac{{\rm Vol}(G)}{(2\pi)^{{\rm
        dim}(G)}}\right)^{2h-2} \sum_{\mu\in P_{++}}\,(\dim\,
V_{\mu})^{2-2h}\,e^{-\sum_{k=1}^{\infty}\,t_k\,p_{k}(\mu+\rho)},}
where $p_{k}\in \IR[\Fh^*]^W$. Note that  $p_{k}(\mu+\rho)$
are equal to the  eigenvalues of particular
combinations of the  (higher) Casimir operator $c_k$ acting on $V_{\mu}$.
For example in the case of the quadratic Casimir operator we have:
$$
c_2|_{V_{\mu}}=\frac{1}{2}(\mu+2\rho,\mu)=\frac{1}{2}(\mu+\rho,\mu+\rho)-
\frac{1}{2}(\rho,\rho)=\frac{1}{2}p_2(\mu+\rho)-\frac{1}{2}(\rho,\rho)$$

In fact, not only the dimensions of the irreducible
representations show up in the 2d Yang-Mills  theory, but also the
characters of the irreducible representations enter the explicit
expressions
  for correlation functions. To illustrate this let us consider
the correlation function  of the operator
$\CO^{(0)}_k=\tr\,\varphi^k$ inserted at the center of the disk $D$.
Boundary conditions are  defined by
fixing the holonomy of the connection over  the boundary.
The explicit representation is given by:
$$
<\CO^{(0)}_n>_D(x)=\left(\frac{{\rm Vol}(G)}{(2\pi)^{{\rm
        dim}(G)}}\right)^{-1} \sum_{\mu\in P_{++}} p_n(\mu+\rho)\,
e^{-\sum_{k=1}^{\infty}
\,t_k\,p_{k}(\mu+\rho)}\,\,\dim\,V_{\mu}\,\,\Phi_{\mu+\rho}(x),$$
where the holonomy of the boundary connection is  $g=\exp(2\pi i x)\in
H$ and $\Phi_{\mu+\rho}(x)=\Delta_G(e^{2\pi i x})\ch_{\mu}(e^{2\pi i x})$.
In particular for $n=0$ and $t_k=0$ we have: \eqn\DISKO{
Z_{YM}(D)\equiv <1>_D(g)=\left(\frac{{\rm Vol}(G)}{(2\pi)^{{\rm
        dim}(G)}}\right)^{-1}\sum_{\mu \in P_{++}}
\,\dim\,V_{\mu}\,\,\Phi_{\mu+\rho}(g)=\delta^{(G)}_e(g),}
where we lift the expression to ${\rm Ad}_G$-invariant functions on
$G$. The delta-function $\delta_e^{G)}(g)$ on the group $G$ has a support
at the unit element $e\in G$ and is a vacuum  wave function
corresponding to a disk. Let us remark that $\delta_e^{(G)}(g)$ can be
considered as a character of the regular
representation of $G$ and \DISKO\ is a decomposition of the regular
representation over the right action of the group $G$.
 Note that the factor $\dim\,V_{\mu}$ enters in the power of the Euler
 characteristic of the disk $\chi(D)=1$, naturally generalizing
the representation \REGULPART\ for the compact surface.

Finally note that \DISKO\  is obviously compatible with \PARTFUNC\
 for $h=0$ if we represent a sphere as glued from two disks: \eqn\GLU{
Z_{YM}(\Sigma_0)=\int_G dg  <1>_D(g)\,<1>_D(g^{-1})=}$$=
\left(\frac{{\rm Vol}(G)}{(2\pi)^{{\rm dim}(G)}}\right)^{-2}
\sum_{\mu\in
P_{++}}\,(\dim\,V_{\mu})^2\,\,\,e^{-\sum_{k=1}^{\infty}\,t_k\,p_{k}(\mu+\rho)}.$$

\subsec{One-dimensional reduction}

Consider the dimensional reduction of  the two-dimensional Yang-Mills
theory to one dimension (quantum mechanics).  We have for the partition
function:
\eqn\FIntQM{
Z_{QM}(\Gamma)=\frac{1}{\Vol(\CG_{\Gamma})}\int \,
D\varphi Da Db D\eta D\zeta\,\,
e^{\frac{1}{2}\int_{\Gamma}\, dt\,(i\Tr\,\varphi\,(\pr_t
  a+[b,a])+\eta\,\zeta
-g_{YM}^2\Tr\,\varphi^2\,{\rm vol}_{\Gamma}) },}
where $\Gamma$ is a trivalent graph supplied with the volume form
${\rm vol}_{\Gamma}$ on its edges,  $\Vol(\CG_{\Gamma})$
is a volume of the gauge group $\CG_{\Sigma}={\rm Map}(\Gamma,G)$.
Here $(a,b)$ and $(\eta,\zeta)$ are 1d
reductions of two-dimensional fields $(A_{\sigma},A_t)$ and
$(\psi_{\sigma}, \psi_t)$. The sewing conditions for the fields on
different edges of the graph are chosen in such way that
 the gauge invariance of \FIntQM\ holds.
 As in the case of the Yang-Mills  theory we
 will be mostly interested in (the generalizations of) the
theory with $g_{YM}^2=0$.

The path integral \FIntQM\ is invariant under the action of the odd
and even vector fields:
\eqn\BRSTYMQM{
Q\,a=i\eta,\,\,\,
\,\,\,\,\,\,\,Q\,b=i\zeta,\,\,\,\,\,\,\,\,\,Q\,\eta=-[a,\varphi],\,\,\,\,\,\,\,
\,\,\,Q\,\zeta=-(\pr_t\varphi+[b,\varphi]),\,\,\,\,\,\,\,\,\,\,Q\,\varphi=0,}
$$
\CL_{\varphi}\,a=-[\varphi,
a],\,\,\,\CL_{\varphi}\,b=-(\pr_t\varphi+[\varphi,b]),
\,\,\,\,\CL_{\varphi}
\,\eta=-[\varphi,\eta],\,\,\,\CL_{\varphi}\,\zeta=-[\varphi,\zeta],
\,\,\,\,\CL_{\varphi}\,\varphi=0,$$
such that $Q^2=i\CL_{\varphi}$.

Consider the simplest case of the Yang-Mills theory for $G=U(N)$
on the circle $\Gamma=S^1$. The bosonic part of the action has the form:
\eqn\actoneQM{
S(\varphi,a,b)=\frac{1}{2}\int_{S^1}\, dt \, \Tr\,(\varphi \pr_t a+
b\,[a,\varphi]).}
Using the invariance with respect to the gauge transformations
 generated by the first-class constraint:
  \eqn\constrQM{
[a,\varphi]=0,}
the phase space of the theory can be reduced to the finite-dimensional
space. The  (open part of the)  phase space is given by:
 \eqn\PSopenQM{
\CM^{(0)}=(T^*\Fh_0)/W,}
where $\Fh_0=\Fh\cap \Fg^{reg}$ is an intersection of the Cartan
subalgebra  $\Fh\subset \Fg$ with the subset $\Fg^{reg}$ of regular
elements of $\Fg$ and $W$ - is Weyl group.  For $G=U(N)$
the set $\Fh/\Fh_0=\cup_{j<k} \{x_j=x_k\}$ is the   main diagonal.

The Hilbert space of the theory can be realized as a space of  ${\rm
Ad}_G$-invariant  functions on $\Fg$. The paring is defined by the
integration with a
 bi-invariant  normalized Haar  measure:
\eqn\PAIRQM{
  <\Psi_1,\Psi_2>=\int_{\Fg} dy\, \overline{\Psi}_1(y)\,\Psi_2(y).}
On ${\rm Ad}_G$-invariant functions it descends to the integral over
Cartan subalgebra $\Fh$:
\eqn\REDPAIRQM{
<\Psi_1,\Psi_2>=\frac{1}{|W|}\int_{\Fh}
dx\,\Delta_{\Fg}^2(x) \overline{\Psi}_1(x)\,\Psi_2(x),}
where the corresponding Jacobian is given by: \eqn\JacobQM{
\Delta_{\Fg}^2(x)=\prod_{\a \in R^+}\,\a(x)^2.}

The  same set of invariant operators \casimirs\
descends onto $\CM^{(0)}$ and we fix the bases of wave functions by the
conditions \EIGEN\
 where $(\l_1,\cdots,\l_N)$ now take values in a
$rank(\Fg)$-dimensional vector space $P(\IR)=P\otimes \IR$.
Similar to the case of wave-function of the Yang-Mills theory
it is useful  to redefine  wave functions by multiplying
them on:
 \eqn\WeylnomodQM{
\Delta_{\Fg}(x)=\prod_{\a \in R^+} \a(x),}
so that the integration measure becomes flat:
\eqn\FALTMES{<\Psi_1,\Psi_2>=\frac{1}{|W|}\int_{\Fh}
 dx\, \overline{\Phi}_1(x)\,\Phi_2(x),}
where $\Phi_i(x)=\Delta_{\Fg}(x)\Psi_i(x)$.

In the case of the dimensionally reduced
theory  the space of states  does not have a
direct connection with the  representation theory of $G$.
However as a  bases of $W$-skew-invariant eigenfunctions
one can use  the renormalized characters  of the finite-dimensional
 irreducible representations
continued to the arbitrary weights $(\lambda_1,\cdots,\lambda_N)$
in the positive Weyl chamber $P_+(\IR)$.
Explicitly we have:  \eqn\wavefuncQM{
\Phi_{\lambda}(x)=\sum_{w\in W}(-1)^{l(w)}e^{2\pi i w(\lambda)(x)}.}

The partition function of the dimensionally reduced topological
Yang-Mills theory  on a graph  $\Gamma_h$
 can be  expressed as a formal integral over the cone $P_{\IR}^+$:
\eqn\PARTFUNCQM{
Z_{YM}(\Sigma_h)=\int_{\lambda\in P_+(\IR)} d^N\lambda
\,\,\,\,\,d_{\lambda}^{2-2h},}
where:
 \eqn\DIMQM{
d_{\lambda}=\prod_{\a\in R_+}(\lambda,\a),}
is a continuation of the renormalized Weyl expression for
the dimension of the irreducible representation $V_{\lambda}$
from $P_{++}$ to $P_+(\IR)$.

\subsec{$G/G$ gauged Wess-Zumino-Witten model}

The  Yang-Mills theory in two dimensions  allows the
 following nontrivial generalization to the $G/G$ gauged WZW model.
The partition function for the $G/G$ gauged WZW model
on $\Sigma_h$ is given by the following path integral: \eqn\GWZW{
 Z_{GWZW}(\Sigma_h)=\frac{1}{{\rm Vol}(\CG_{\Sigma_h})}
\int Dg\,DA\,D\psi\,\,\,e^{kS(g,A,\psi)},}
$$S(g,A,\psi)=S_{WZW}(g)-\frac{1}{2\pi}\int_{\Sigma_h}d^2z\,
\Tr\,(A_zg^{-1}\apr_{\zb}g+
g\pr_zg^{-1}A_{\zb}+gA_zg^{-1}A_{\zb}-A_zA_{\zb})+$$
$$+\frac{1}{4\pi}\int_{\Sigma_h}\,d^2z\,
\Tr(\psi\wedge \psi),  $$
where $S_{WZW}(g)$ is an action functional for Wess-Zumino-Witten model:
\eqn\WZW{
S_{WZW}=-\frac{1}{8\pi}\int_{\Sigma_h}d^2z\,
\Tr(g^{-1}\pr_zg\cdot g^{-1}\pr_{\zb}g)-i\Gamma(g),}
$$
\Gamma(g)=\frac{1}{12\pi}\int_Bd^3y\,\epsilon^{ijk}\,
\Tr\,g^{-1}\pr_ig\cdot g^{-1}\pr_jg\cdot g^{-1}\pr_kg.$$
Here $k$ is a positive  integer and $\pr B= \Sigma_h$.

The gauged WZW model  allows even and odd
symmetries extending those for the Yang-Mills theory
 \ref\AG{  A.~Gerasimov,   {\it Localization in GWZW and Verlinde
formula}, [arXiv:hep-th/9305090].}:
\eqn\BRSTYMGWZW{
Q\,A=i\psi,\,\,\,\,\,\,Q\,\psi^{(1,0)}=i(A^g)^{(1,0)}-iA^{(1,0)},
\,\,\,\,\,\,\,Q\,\psi^{(0,1)}=-i(A^{g^{-1}})^{(0,1)}+iA^{(0,1)},}
$$Q\,g=0,$$
\eqn\GAUGEYMGWZW{
\CL_{g}\,A^{(1,0)}=(A^g)^{(1,0)}-A^{(1,0)}
\,\,\,\,\,\,\,\CL_g\,A^{(0,1)}_A=-(A^{g^{-1}})^{(0,1)}+A^{(0,1)},}
$$\CL_{g}\,\psi^{(1,0)}=-g\psi^{(1,0)} g^{-1}+\psi^{(1,0)},\,\,\,\,\,\,
\CL_{g}\,\psi^{(0,1)}=g^{-1}\psi^{(0,1)}g-\psi^{(0,1)},\,\,\,\,\CL_{g}\,g=0.$$
 Here $A^g=g^{-1}dg+g^{-1} A g$ is a gauge transformation.
We have the following relation  $Q^2=\CL_{g}$.
It is useful to compare these generators with
those in the pure Yang-Mills theory. Generators
 \GAUGEYM\ and \BRSTYM\ realize infinitesimal symmetries
(from the Lie algebra) of the action
 functional. In contrast  the transformations \GAUGEYMGWZW\
and \BRSTYMGWZW\ realize the finite (from the gauge group) symmetries of
the action.
Note that in the limit $g\to 1+i\epsilon \varphi_0$, $\epsilon \to 0$
\GAUGEYMGWZW\ and \BRSTYMGWZW\  are reduced to \GAUGEYM\
and \BRSTYM.

Consider a deformation of the theory by:
\eqn\DEFGWZW{
  \Delta S=\sum_{\mu\in P_{++}} t_{\mu}\,\int_{\Sigma_h} d^2z
\,\Tr_{V_{\mu}}
g\,\,{\rm vol}_{\Sigma_h}.}
We take $t_{\mu}=0$ for all but finite subset of $P_{++}$
to make the path integral well defined.
Similarly to \PARTFUNC\
the partition function can be represented in the following form:
\eqn\GWZWPARTFUNC{
  Z_{GWZW}(\Sigma_h)=|Z(G)|^{2h-2}(\frac{k+c_v}{4\pi^2})^{\frac{1}{2}{\rm
      dim}\,\CM_G(\Sigma_h)} {\rm
    Vol}_q(G)^{2h-2}\times } $$
\times \sum_{\mu\in P_{++}^k}\,({\rm dim}_q\, V_{\mu})^{2-2h}
\,e^{-\sum_{\mu\in P_{++}^k} t_{\mu}{\rm ch}_{V_{\mu}}(e^{2\pi i
\hat{\l}})}\,\,\,\,\,.$$
where ${\rm  dim}\,\CM_G(\Sigma_h)={\rm dim}\,G (2h-2)$ is the
dimension of the moduli space of flat $G$-bundles on $\Sigma_h$,
$|Z(G)|$ is a dimension of the center of $G$ and:
\eqn\QDIM{
{\rm dim}_q\, V_{\mu}={\rm Tr}_{V_\mu}q^{-\hat{\rho}}=\prod_{\a\in R_+}
\frac{(q^{\frac{1}{2}(\mu+\rho,\a)}-q^{-\frac{1}{2}(\mu+\rho,\a)})}
{(q^{\frac{1}{2}(\rho,\a)}-q^{-\frac{1}{2}(\rho,\a)})},}
\eqn\QVOL{
{\rm Vol}_q(G)=(2\pi)^{{\rm dim}G}(k+c_v)^{-\frac{1}{2}({\rm dim}\,G -{\rm
dim}\,H)}
\prod_{\a\in R_+}\left(q^{\frac{1}{2}(\rho,\a)}
-q^{-\frac{1}{2}(\rho,\a)}\right)^{-1},}
and the sum is over the set $P_{++}^k$ of
integrable representations of the affine group $\widehat{LG}_k$
 at the level $k$. The same set also enumerates   irreducible
representations
of  $U_q({\Fg})$ for $q=\exp(2\pi i/(k+c_v))$.
We define $\exp(2\pi i \hat{\l})=\exp (2\pi i \sum_j \l_j\,e_j)\in H$ and
$
{\rm ch}_{V_{\mu}}(e^{2\pi i \hat{\l}})$ is a character of the element
$\exp(2\pi i \hat{\l})$ taken in
the representation $V_{\mu}$. The expressions ${\rm dim}_q\, V_{\mu}$
are known as  quantum dimensions of the
representations of the quantum group. For example in the case of
 $\Fg={\frak  gl}_N$ we have:
\eqn\QDIMGLN{
{\rm dim}_q\,
V_{\mu}=\prod^N_{i<j}\frac{(q^{\frac{1}{2}(\mu_i-\mu_j+j-i)}-
q^{\frac{1}{2}(\mu_j-\mu_i+i-j)})}{(q^{\frac{1}{2}(j-i)}
-q^{\frac{1}{2}(i-j)})}.}
and $\mu_1\geq \mu_2\geq \cdots \geq \mu_N$, $\mu_i\in \IZ_+$.
The $q$-analog of the character is given by:
\eqn\QCHAR{
\Psi_{\mu}(x)=\ch_{\mu}q^{-\rho+x}.}
and corresponding $q$-generalizations of \DISKO\ holds.
The representation \GWZWPARTFUNC\
 for the partition function of the gauged WZW model
can be derived using  the cohomological localization technique
\AG\   (see also \ref\BT{
  M.~Blau and G.~Thompson,
   {\it Derivation of the Verlinde formula from Chern-Simons theory and
the G/G
  model},   Nucl.\ Phys.\ B {\bf 408}, 345 (1993)
  [arXiv:hep-th/9305010].} for slightly different approach).
The choice of the regularization scheme leading to a
particular normalization of the partition function \GWZWPARTFUNC\ is
compatible with the interpretation of  \GWZWPARTFUNC\ as  number of
the conformal blocks in WZW theory on $\Sigma_h$. This interpretation
follows from
the fact that the partition  function in $G/G$ gauged WZW model
coincides with partition function
in 3d Chern-Simons theory for three-dimensional  manifold $\Sigma_h \times
S^1$.

\newsec{Topological Yang-Mills-Higgs theory }

In \mns\ a  two dimensional gauge
theory was proposed such that the space of classical solutions
of the theory on a Riemann surface $\Sigma_h$ is closely related with
the cotangent space to the space of solutions of the dimensionally
reduced (to 2d)  four-dimensional self-dual Yang-Mills   equations studied by  Hitchin
\Hitchin. Below we present a way to calculate the  partition function of the theory
following \mns. Let us given a principal $G$-bundle $P_G$ over a Riemann surface
$\Sigma$ supplied with a complex structure. Then  we have an associated
vector bundle ${\rm ad}_{\Fg}=(P_G\times \Fg)/G$ with
the fiber $\Fg=Lie(G)$ supplied with
a coadjoint action of $G$.  Consider the
pairs $(A,\Phi)$ where  $A$ is a connection on $P_G$ and $\Phi$ is a
one-form taking values in ${\rm ad}_{\Fg}$. Then Hitchin equations are
given by:
\eqn\HEQUAT{
F(A)-\Phi \wedge
\Phi=0,\,\,\,\,\,\,\nabla_A^{(1,0)}\Phi^{(0,1)}=0,\,\,\,\,\,\,
 \nabla_A^{(0,1)}\Phi^{(1,0)}=0.}
The space of the solutions has a natural hyperk\"{a}hler structure and
admits  compatible $U(1)$ action. The correlation functions in the theory
introduced in \mns\ can be described by the integrals of the products
of $U(1)\times G$-equivariant cohomology classes over the moduli space
of solutions of \HEQUAT.   Note that the $U(1)$-equivariance  makes the
path integral well-defined.

The  field content of the theory introduced in \mns\ can be described
as follows. In addition to the  triplet $(A,\psi_A,\varphi_0)$
of the  topological  Yang-Mills  theory one has:
\eqn\FCONT{
(\Phi,\psi_{\Phi}):
\,\,\,\,\,\,\,\,\,\,\,\Phi\in
\CA^1(\Sigma, {\rm ad}_{\Fg}),\,\,\,\,\,\,\,\,\,
\psi_{\Phi}\in \CA^1(\Sigma, {\rm ad}_{\Fg})
}
\eqn\FCONTone{
(\varphi_{\pm},\chi_{\pm}):\,\,\,\,\,\,\,\,\,\,\varphi_{\pm}\in
\CA^0(\Sigma, {\rm ad}_{\Fg}),
\,\,\,\,\,\,\,\,\,\,\,\chi_{\pm}\in \CA^0(\Sigma, {\rm ad}_{\Fg})}
where $\Phi$, $\varphi_{\pm}$ are even  and $\psi_{\Phi}$,
$\chi_{\pm}$ are odd fields. We will use also another notations
  $\varphi_{\pm}=\varphi_1\pm i\varphi_2$.

The theory is described by the following path integral:
 \eqn\YMH{
Z_{YMH}(\Sigma_h)=\frac{1}{{\rm Vol}(\CG_{\Sigma_h})}
\int\, D\varphi_0\,D\varphi_{\pm}\,DA\,D\Phi\,
D\psi_A \,D\psi_{\Phi} D\chi_{\pm}\,\, e^{S(\varphi_0,\varphi_{\pm}
,A,\Phi,\psi_A,\psi_{\Phi},\chi_{\pm})},}
where $S=S_0+S_1$ with:
\eqn\ACTBOS{\eqalign{
S_0(\varphi_0,\varphi_{\pm},A,\Phi,\psi_A, & \psi_{\Phi},\chi_{\pm})=
\frac{1}{2\pi}\int_{\Sigma_h} d^2z\,\Tr(i\varphi_0\,(F(A)-
\Phi\wedge \Phi)-c\Phi\wedge *\Phi)+\cr
+ &
\varphi_+\nabla^{(1,0)}_A\,\Phi^{(0,1)}+\varphi_-\nabla^{(0,1)}_A\,\Phi^{(1,0)}}}
and
\eqn\ACTFERM{\eqalign{
S_1(\varphi_0,\varphi_{\pm},A,\Phi,\psi_A,\psi_{\Phi},\chi_{\pm})=
\frac{1}{2\pi}\int_{\Sigma_h} d^2z\,\Tr(\frac{1}{2}\psi_{A}\wedge
\psi_{A}+
\frac{1}{2}\psi_{\Phi}\wedge \psi_{\Phi}+\cr
+\chi_+[\psi^{(1,0)}_A,\Phi^{(0,1)}]+\chi_-[\psi^{(0,1)}_A,\Phi^{(1,0)}]
+\chi_+\nabla^{(1,0)}_A \psi^{(0,1)}_{\Phi}+\chi_-\nabla^{(0,1)}_A
\psi^{(1,0)}_{\Phi})}}
where the decompositions $\Phi=\Phi^{(1,0)}+\Phi^{(0,1)}$ and
$\psi_{\Phi}=\psi^{(1,0)}_{\Phi}
+\psi^{(0,1)}_{\Phi}$ correspond to the  decomposition of
the space of one-forms $\CA^1(\Sigma_h)
=\CA^{(1,0)}(\Sigma_h)\oplus \CA^{(0,1)}(\Sigma_h)$
defined in terms of  a fixed complex structure on $\Sigma_h$.

 The theory is invariant under the action of  the  following even vector
field:
\eqn\UONEACT{
\CL_v\Phi^{(1,0)}=+ \Phi^{(1,0)}, \,\,\,\,\,
\CL_v\Phi^{(0,1)}=-\Phi^{(1,0)},\,\,\,\,\,
\CL_v\psi^{(1,0)}_{\Phi}=+  \psi^{(1,0)}_{\Phi},}
$$\CL_v\psi^{(0,1)}_{\Phi}=- \psi^{(0,1)}_{\Phi}\,\,\,\,\,
\CL_v\varphi_{\pm}=\mp\varphi_{\pm}, \,\,\,\,\,
\CL_v\chi_{\pm}=\pm \chi_{\pm},\,\,\,\,\,\,\,$$
\eqn\GAUGEYM{
\CL_{\varphi_0}\,A=-\nabla_A\varphi_0,\,\,\,\,
\CL_{\varphi_0}\,\psi_A=-[\varphi_0,\psi_A],
\,\,\,\,\CL_{\varphi_0}\,\Phi=-[\varphi_0,\Phi],\,\,\,
\CL_{\varphi_0}\,\psi_{\Phi}=-[\varphi_0,\psi_{\Phi}],}
$$\CL_{\varphi_0}\,\varphi_0=0,\,\,\,\,\,\,\CL_{\varphi_0}\,\varphi_{\pm}=
-[\varphi_0,\varphi_{\pm}],\,\,\,\,\,\,\,\CL_{\varphi_0}\,\chi_{\pm}=
-[\varphi_0,\chi_{\pm}],   $$
and an odd vector field generated by the BRST operator:
$$
QA=i\psi_A,\,\,\,\, Q\psi_A=-\nabla_A\varphi_0,\,\,\,\,Q\varphi_0=0,$$
\eqn\BRST{
Q\Phi=i\psi_{\Phi},\,\,\,\,Q\psi^{(1,0)}_{\Phi}=-[\varphi_0,\Phi^{(1,0)}]+c\Phi^{(1,0)},
\,\,\,\,\,Q\psi^{(0,1)}_{\Phi}=-[\varphi_0,\Phi^{(0,1)}]-c\Phi^{(0,1)},}
$$Q\chi_{\pm}=i\varphi_{\pm},\,\,\,\,Q\varphi_{\pm}=-[\varphi_0,\chi_{\pm}]\pm
c\chi_{\pm}.$$
We have $Q^2=i\CL_{\varphi_0}+c\CL_v$ and $Q$ can be considered as a
BRST operator on the space of $\CL_{\varphi_0}$ and $\CL_v$-invariant
functionals. The action functional of the topological Yang-Mills-Higgs
theory can be represented as a sum of the action functional of the
topological  pure Yang-Mills theory (written in terms of  fields $\phi_0,
A, \psi_A$) and an additional part which can be
represented as a $Q$-anti-commutator:
\eqn\ACTONQCOM{
S_{YMH}=S_{YM}+[Q,\int_{\Sigma_h}\,d^2z \,\Tr\,(\frac{1}{2}
 \Phi \wedge \psi_{\Phi}+\varphi_+
 \nabla^{(1,0)}_A\Phi^{(0,1)}+\varphi_-
 \nabla^{(0,1)}_A\Phi^{(1,0)})]_+.}
  The theory given by \YMH\ is a  quantum field theory whose
correlation functions are given by the intersections pairings of the
equivariant cohomology classes on the moduli spaces
of Higgs bundles.

To simplify the calculations
it is useful to consider  more general  action given by:
\eqn\ACTTWOPAR{
S_{YMH}=S_{YM}+[Q,\int_{\Sigma_h}\,d^2z \,\, \Tr(\frac{1}{2}
 \Phi\wedge \psi_{\Phi}+}$$+\tau_1\,(\varphi_+
 \nabla^{(1,0)}_A\Phi^{(0,1)}+\varphi_-
\nabla^{(0,1)}_A\Phi^{(1,0)})+\tau_2(\chi_+\varphi_-+\chi_-\varphi_+)\,{\rm
  vol}_{\Sigma_h})].$$
Cohomological localization of the functional integral
takes the simplest form for   $\tau_1=0$, $\tau_2\neq 0$.
Note that it is not obvious that the theory for $\tau_1=0$, $\tau_2\neq 0$
is equivalent to that for $\tau_1\neq 0$, $\tau_2=0$.
However taking into account that the action functionals in these two
 cases differ on the equivaraintly exact form and
for $c\neq 0$ the space of fields is
essentially compact one can expect that the theories are equivalent.

For  $\tau_1=0$ the path integrals over $\Phi$, $\varphi_{\pm}$ and
$\chi_{\pm}$ is quadratic. Thus we have for the partition function
the following formal representation:
\eqn\NONLOCOB{
Z_{YMH}(\Sigma_h)=}
$$=\frac{1}{{\rm Vol}(\CG_{\Sigma_h})}
\int DA\,D\varphi_0\,D\psi_A
\,\,e^{\frac{1}{2\pi}\int_{\Sigma_h}d^2z\,\Tr\,(
i\varphi_0 F(A)+\frac{1}{2}\psi_A\wedge \psi_A)}\,\,
{\rm Sdet}_V(ad_{\varphi_0}+ic),$$
where  the  super-determinant is taken over the super-space:
\eqn\supersp{
V=V_{even}\oplus V_{odd}=\CA^0(\Sigma_h,{\rm ad}\Fg)\oplus
\CA^{(1,0)}(\Sigma_h,{\rm ad}\Fg).}
and should be properly understood using
a regularization compatible with $Q$-symmetry of the path integral
(e.g. $\tau_1\neq 0$).
Thus the Yang-Mills-Higgs theory can be considered as a
 pure Yang-Mills theory deformed by a non-local gauge invariant
observable.

Let us stress  that there are two interesting limiting cases
 $c \rightarrow \infty$ and $c \rightarrow 0$
for the theory \YMH, obvious from the definition of the corresponding
Lagrangian.
Note that the dependence on $c$ is through the mass term for the field
$\Phi$ in \ACTBOS\ and \ACTFERM. Thus in the limit $c \rightarrow \infty$
the  field $\Phi$ (and corresponding fermions) drops
out and we get the 2d Yang-Mills theory with a compact group $G$.
 Therefore, in this limit we have to recover the known answers from
2d Yang-Mills theory. On the other hand for $c=0$
 the topological Yang-Mills-Higgs theory for group $G$
is equivalent to  2d topological Yang-Mills theory for
complex group $G^c$. This might be considered as a manifestation of the
general
relation between the hyperk\"{a}hler quotient over a compact group
and a  K\"{a}hler quotient over its complexification.
In the case of  \YMH\ the gauge symmetry group is
$\CG={\rm Map}(\Sigma_h,G)$ while in the complexified Yang-Mills theory
one would have $\CG^c={\rm Map}(\Sigma_h,G^c)$ as a gauge group.
Thus in the limit $c\to 0$ one might expect the relation
with  the representation theory of the complexified group $G^c$.

Let us demonstrate the relation of $c=0$ Yang-Mills-Higgs theory
with the  Yang-Mills theory for complexified gauge group explicitly.
The Feynman path integral representation for Yang-Mills-Higgs
theory at $c=0$ in \YMH\ can be considered as a result of a partial gauge
fixing of the
symmetry in the complex Yang-Mills theory. Consider a  complex gauge
field:
\eqn\COMCON{\nabla_{\CA}=d+A^c=d+A+i\Phi,}
 where $A$ and  $\Phi$ are skew-hermitian one forms.
The corresponding curvature is naturally
decomposed into  the skew-hermitian  and hermitian parts:
\eqn\CURV{
F(A^c)=(dA+A\wedge A-\Phi\wedge \Phi)+i(d\Phi+A\wedge \Phi).}
Define two-dimensional topological Yang-Mills theory for the complex
group $G^c$ as:
\eqn\CYM{
Z_{YM^c}(\Sigma_h)=\frac{1}{{\rm Vol}(\CG^c_{\Sigma_h})}
\int\,D\varphi_0\,D\varphi_-\,DA\,D\Phi\,\,e^{S_{YM^c}(\varphi_{\pm},
A,\Phi)},}
where:
\eqn\CACT{
S_{YM^c}(\varphi_0,\varphi_1,A,\Phi)=
\frac{1}{2\pi}\int_{\Sigma_h}
d^2z\,\Tr\,(\varphi_c\,F(A^c)+\overline{\varphi}_c\,F(\overline{A^c}))=}
$$
=\frac{1}{2\pi}\int_{\Sigma_h}\,d^2z\,\,\Tr(\varphi_0(dA+A\wedge
A-\Phi\wedge \Phi)+
\varphi_1(d\Phi+A\wedge \Phi)),$$
and $\varphi_c=\varphi_0+i\varphi_1$.
This theory has an infinitesimal  gauge symmetry:
\eqn\CGTR{
A\rightarrow d\epsilon+[\epsilon,A]-[\eta,\Phi],\,\,\,\,\,\,\,\,
\Phi\rightarrow d\eta+[\eta,A]+[\epsilon,\Phi],}
with the gauge parameter $(\epsilon+i\eta)\in \CA^0(\Sigma_h,\Fg^c)$.
To make contact with \YMH\ one should  partially fix the gauge freedom
generated by the $\eta$-dependent part of \CGTR\ by adding:
\eqn\CGFIX{
\Delta S=\int_{\Sigma_h} d^2z\,
\Tr(\varphi_2\,\nabla_A(*\Phi))+}$$+\int_{\Sigma_h}d^2z\, \Tr\,
\frac{1}{2}
(\nabla^{1,0}_A\chi_{+}+i[\Phi^{(1,0)},\chi_{+}])
(\nabla^{0,1}_A\chi_{-}+i[\Phi^{(0,1)},\chi_-]),$$
where last term is the ghost-antighost contribution.
Note that this term is invariant with respect to $\epsilon$-symmetry
 if the  field $\varphi_-$ takes values in the coadjoint
 representation of the group. Taking $c=0$ and integrating  over
$\psi_{A}$ and $\psi_{\Phi}$ in \YMH\ one can see that the theory \YMH\ is
equivalent to this, partially gauge fixed, complex Yang-Mills theory.

The partition function \YMH\ of the Yang-Mills-Higgs theory on a compact
Riemann
surface can be calculated using the standard methods of the
cohomological localization \Wtwo.
As in the case of Yang-Mills theory we consider the deformation of
 the action of the theory:
\eqn\DEFACTNone{
\Delta S_{YMH}=-\sum_{k=1}^{\infty}\,t_k\,\,\int_{\Sigma_h}d^2z \,\,
\Tr\,\varphi_0^{k}\,\,{\rm vol}_{\Sigma_h}.}
where we impose the condition that $t_k$ for all but finite set of
indexes.

Path integral with the action \ACTTWOPAR\ at $\tau_1=0$ and $\tau_2=1$ is
easily reduced to the integral over abelian gauge fields. The
  contribution of the additional nonlocal observable in
 \NONLOCOB\ can be calculated  as follows.
 The purely bosonic part of the nonlocal observable
 after reduction to abelian fields can be
easily evaluated using any suitable regularization (i.e. zeta
function regularization) and result is the change of the bosonic
part of the abelian action $\int d^2z(\varphi_0)_iF^i(A)$ by:
\eqn\SHITFS{\eqalign{ \Delta S=&\int_{\Sigma_h}\,d^2z
\,\sum_{i,j=1}^N \log\left(\frac{(\varphi_0)_i-(\varphi_0)_j+ic}{
    (\varphi_0)_i-(\varphi_0)_j-ic}
\right) F(A)^i \cr
& +\frac{1}{2}\int_{\Sigma_h}\,d^2z
\,\sum_{i,j=1}^N \log((\varphi_0)_i-(\varphi_0)_j+ic)R^{(2)}\sqrt{g};}}
where $F(A)^i$ is $i$-th component of the curvature
of the abelian connection $A$ and $R^{(2)}(g)$ is curvature on
$\Sigma_h$ for 2d metric $g$
used to regularize non-local observable.  We will use the notation
$(\varphi_0)_i=\l_i$ in remaining part of the paper. This leads to
the unique $Q$-closed
completion. The completion of the term in \SHITFS\ containing the
curvature of the gauge field  is given by the
two-observable $\CO^{(2)}_f$ corresponding to the descendent of the
following function on the Cartan subalgebra isomorphic to $\IR^N$:
 \eqn\INVF{
f(diag(\l_1,\cdots ,\l_N)=\sum_{k,j=1}^N \int_0^{\l_j-\l_k}\,{\rm
arctg}\, \l/c\,d\l. }
Thus, the abelianized action is defined by two-observable descending
from:
\eqn\yy{ I(\l)=\sum_{j=1}^N (\frac{1}{2}
\l_i^2-2\pi  n_j\l_j)+\sum_{k,j=1}^N \int_0^{\l_j-\l_k}\,{\rm
arctg}\,\, \l/c\,d\l,}
according the formula \OBTWO.
On the other hand the term containing the metric curvature $R$
in \SHITFS\ is $Q$-closed and thus does not need any completion.
It can be considered as an integral of the zero observable:
\eqn\ZEROCURV{
\CO^{(0)}=\sum_{i,j=1}^N \log((\varphi_0)_i-(\varphi_0)_j+ic)}
over $\Sigma_h$ weighted by the half of the metric curvature.
Note that the function $I(\l)$ plays important role in Nonlinear
Schr\"{o}dinger
theory which we explain in next section.

After integrating out the fermionic
partners of abelian connection $A$ the standard localization
procedure leads to following final finite-dimensional integral
representation for the
the partition function \mns: \eqn\SDHsum{
Z_{YMH}(\Sigma_h)=\frac{e^{(1-h)a(c)}}{|W|} \int_{\IR^N} d^N\l\,\,\,
\mu(\l)^{h} \,\sum_{(n_1,\cdots,n_N)\in \IZ^N} e^{2\pi i
\sum_{m=1}^N\l_m n_m}\times }
$$\times  \prod_{k\neq j}(\l_k-\l_j)^{n_k-n_j+1-h}
\,\,\prod_{k,j}(\l_k-\l_j-ic)^{n_k-n_j+1-h}\,e^{-\sum_{k=1}^{\infty}
\,t_k\,\,p_{k}(\l)} ,$$
where
\eqn\MEASURE{
\mu(\l)=\det \|\frac{\pr^2 I(\l)}{\pr \l_i \,\pr \l_j}\|,}
and $p_{k}(\l)$ are $S_N$-invariant  polynomial functions of degree
$k$ on $\IR^N$ and $a(c)$ is a $h$-independent constant
defined by the appropriate choice of the regularization  of the functional
integral.
 One can write the $n_i$-dependent parts of the products in
  \SDHsum\ as the exponent of the sum:
\eqn\SDHsumon{ Z_{YMH}(\Sigma_h)=\frac{e^{(1-h)a(c)}}{|W|}
\int_{\IR^N} d^N\l\,\, \mu(\l)^{h}\, \sum_{(n_1,\cdots,n_N)\in \IZ^N}
e^{2\pi i \sum_j n_j\alpha_j(\lambda)}}
$$\times  \prod_{k\neq j}(\l_k-\l_j)^{1-h}
\,\,\prod_{k,j}(\l_k-\l_j-ic)^{1-h}\,e^{-\sum_{k=1}^{\infty}
\,t_k\,\,p_{k}(\l)} ,$$
with notation:
\eqn\yang{e^{2\pi i \alpha_j(\lambda)}=\CF_j(\l)\equiv
e^{2\pi i \l_j}\prod_{k\neq j}\frac{\l_k-\l_j-ic}{\l_k-\l_j+ic}}
 After taking the sum over $(n_1,\cdots,n_N)\in \IZ^N$  using:
\eqn\correc{\eqalign{\mu(\l) \sum_{(n_1,\cdots,n_N)\in \IZ^N} & e^{2\pi i
\sum_j n_j\alpha_i(\lambda)}=\mu(\l)
\sum_{(m_1,\cdots,m_N)\in \IZ^N}\prod_j\delta(\alpha_j(\l)-m_j)\cr
&
=\sum_{(\l^*_1,\cdots,\l^*_N)\in \CR_N}\prod_j\delta(\l_j-\l_j^*)}}
(see definition of $\CR_N$ below)  and integral over
$(\l_1,\cdots ,\l_N)\in \IR^N$
  we see  that only $\alpha_j(\lambda) \in \IZ$,
or the same - $\CF_j(\l)=1$, contribute to the partition
function which now can be written in the form similar to \PARTFUNC,
\GWZWPARTFUNC\  for $\Fg={\frak u}_N$:
\eqn\SDHPART{
Z_{YMH}(\Sigma_h)=e^{(1-h)a(c)}\sum_{\l\in \CR_N} \,
D_{\l}^{2-2h}\,e^{-\sum_{k=1}^{\infty}
\,t_k\,\,p_{k}(\l)},}
where:
\eqn\SDHDIM{
D_{\l}=\mu(\l)^{-1/2}\prod_{i<j}(\l_i-\l_j)
(c^2+(\l_i-\l_j)^2)^{1/2},}
and the   $\CR_N$ in \correc\ and \SDHPART\ denotes a  set of the
solutions
of the Bethe Ansatz  equations $\CF_j(\l)=1$:
\eqn\BA{
e^{2\pi i\l_j}\prod_{k\neq
  j}\frac{\l_k-\l_j+ic}{\l_k-\l_j-ic}=1,\qquad\,\,\,\,\,\,\,\,\,\,\,\,
k=1,\cdots
,N,}
for the $N$-particle  sector of the quantum theory of
 Nonlinear Schr\"{o}dinger  equation
 (see e.g. \Ga, \IKS). Note that the sum in \SDHPART\ is taken over the
 classes of the solutions  up to action of the
symmetric group on $\l_i$. This set can be  enumerated by the
 multiplets of the integer numbers $(p_1,\cdots, p_N)\in \IZ^N$ such
 that $p_1\geq  p_2\geq \cdots \geq
 p_N$, $\,p_i\in \IZ$.  Thus, the sum in \SDHPART\
is the sum over the same set of partitions
 as in 2d Yang-Mills theory. The structure of the representation
\SDHPART\ for the partition function of Yang-Mills-Higgs theory is
very similar to the analogous representation \PARTFUNC, \GWZWPARTFUNC\
of the partition functions for Yang-Mills and gauged
Wess-Zumino-Witten theories.

\subsec{Reduction of  Yang-Mills-Higgs theory  to one dimension}

Let us use the following notations for the one-dimensional reduction
 of the non-scalar fields entering the description of Yang-Mills-Higgs
theory:
\eqn\REDYMH{
A\to (a,b),\,\,\,\,\Phi \to (\phi,\rho),\,\,\,\,\psi_A\to
(\eta_a,\zeta_b), \,\,\,\, \psi_{\Phi}\to (\eta_{\Phi},\zeta_{\Phi})
}
Reduction of topological Yang-Mills-Higgs
 theory to one dimension (Yang-Mills-Higgs Quantum Mechanics)
is described by the following
path  integral:
$$Z_{YMHQM}(\Gamma_h)=\int\, D(\varphi_0,\varphi_{\pm}
,a,b,\phi,\rho,\eta_a,\zeta_b,\eta_{\phi},\zeta_{\rho},\chi_{\pm})\,\,\,\,
\, e^{S(\varphi_0,\varphi_{\pm}
,a,b,\phi,\rho,\eta_a,\zeta_b,\eta_{\phi},\zeta_{\rho},\chi_{\pm})},$$
where  the action is given by $S=S_0+S_1$:
 \eqn\ACTBOS{
S_0=\frac{1}{4\pi}
\int dt\,\Tr\,(i\varphi_0\,(\pr_t a+[b,a]+[\phi,\rho])
-c(\phi^2+\rho^2)+}
$$+
\varphi_1(\pr_t\phi+[b,\phi]-[a,\rho])+\varphi_2(\pr_t\rho+[b,\rho]+[a,\phi])),$$
\eqn\ACTFERM{
S_1=\frac{1}{2\pi}\int dt\,\Tr\,(\frac{1}{2}\eta_{a}\zeta_{b}+
\frac{1}{2}\eta_{\phi}\zeta_{\rho}+
\chi_1([\eta_a,\phi]+[\zeta_b,\rho])+\chi_2([\zeta_{b},\phi]-[\eta_{a},\rho])+}$$
+\chi_1(\pr_t\eta_{\Phi}+[b,\eta_{\Phi}]-[a,\zeta_{\Phi}])+
\chi_2(\pr_t\zeta_{\Phi}+[b,\zeta_{\Phi}]+[a,\eta_{\Phi}])).$$
 The  theory is invariant under the action of  the  vector fields:
$$
\CL_v\phi=+\rho, \,\,\,\,\,\CL_v\rho=-\phi,\,\,\,\,\,
\CL_v\eta_{\Phi}=-\zeta_{\Phi},$$
$$\CL_v\,\zeta_{\Phi}=+ \eta_{\Phi}\,\,\,\,\,
\CL_v\,\varphi_{\pm}=\mp\varphi_{\pm}, \,\,\,\,\,
\CL_v\,\chi_{\pm}=\pm \chi_{\pm},\,\,\,\,\,\,\,$$
\eqn\GAUGEYMQM{
\CL_{\varphi_0}\,a=-[a,\varphi_0,],\,\,
\CL_{\varphi_0}\,b=-(\pr_t\varphi_0+[b,\varphi_0,]),\,\,\,\,
\CL_{\varphi_0}\,\eta_a=-[\varphi_0,\eta_a],\,\,\,\,\,
\CL_{\varphi_0}\,\zeta_b=-[\varphi_0,\zeta_b],}
$$\CL_{\varphi_0}\,\phi=-[\varphi_0,\phi],\,\,\,
\CL_{\varphi_0}\,\rho=-[\varphi_0,\rho],\,\,\,\,
\CL_{\varphi_0}\,\eta_{\Phi}=-[\varphi_0,\eta_{\Phi}],\,\,\,\,\,
\CL_{\varphi_0}\,\zeta_{\Phi}=-[\varphi_0,\zeta_{\Phi}],$$
$$\CL_{\varphi_0}\,\varphi_0=0,\,\,\,\,\,\,\CL_{\varphi_0}\,\chi_{\pm}=
-[\varphi_0,\chi_{\pm}],   $$
and a fermionic symmetry generated by the BRST operator:
\eqn\BRST{
Q a=i\eta_{a},\,\,\,Q b=i\zeta_{b},\, Q\eta_a=-[a,\varphi_0],\,\,
Q\zeta_b=-(\pr_t b+[b,\varphi_0]),\,\,\,\,Q\varphi_0=0,}
\eqn\BRSTone{
Q\phi=i\eta_{\Phi},\,\,\, Q\rho=i\zeta_{\Phi},\,\,\,\,
Q\,\eta_{\Phi}=-[\varphi_0,\phi]+c\rho,
\,\,\,\,
Q\,\zeta_{\Phi}=-[\varphi_0,\rho]-c\phi,}
\eqn\BRSTtwo{
Q\chi_{\pm}=i\varphi_{\pm},\,\,\,\,Q\varphi_{\pm}=-[\varphi_0,\chi_{\pm}]\pm
c\chi_{\pm}.}
We have $Q^2=i\CL_{\varphi_0}+c\CL_v$ and $Q$ can be considered as a
BRST operator on the space of $\CL_{\varphi_0}$ and $\CL_v$-invariant
functionals.

The partition function on the graph $\Gamma_h$ for $\Fg={\frak u}_N$
 after localization is given by: $$
Z_{YMHQM}(\Gamma_h)=\frac{e^{(1-h)a(c)}}{|W|}\int_{\IR^N/S_N}\,d^N\l
 \,\,\,\prod_{k\neq j}(\l_k-\l_j)^{1-h}\,\,
\prod_{k\neq j}(\l_k-\l_j-ic)^{1-h}=$$
\eqn\SDHsumone{=
\frac{e^{(1-h)a(c)}}{|W|}\int_{\IR^N/S_N} d^N\l\,\,
D_{\l}^{2-2h},}
where:
\eqn\SDHDIMo{
D_{\l}=\prod_{i<j}(\l_i-\l_j)
(c^2+(\l_i-\l_j)^2)^{1/2}.}
In contrast with the two-dimensional case we have integral over
$\IR^N/S_N$ instead of the sum over the solutions of Bethe Ansatz
equations. Note also that the factor $\mu(\l)$ is a constant for
  the dimensionally reduced theory and is included into the proper
  normalization of the partition function.  If we compare
\SDHsumon\ and \SDHsumone\ we see that the only difference is that in
\SDHsumon\ we have additional insertion under integral over $\lambda$'s of
the sum
over integers $(n_1,\cdots,n_N)$  with the exponential factor that reduces
the integral to the sum over the zeros of the exponent,
$\alpha_i(\lambda)$, which is the
same as a reduction to those $\lambda$'s that solve Bethe Ansatz equation.
 This simple fact will be important
later in computation of wave functions for Yang-Mills-Higgs theory.

Obvious similarities between \SDHPART\ and  \PARTFUNC,
\GWZWPARTFUNC\ suggests that there should be the full analog of
the results discussed in the Section 2, including the
interpretation of $D_{\l}$ as a (formal) dimension of the
representation of some algebraic structure  together with the
identification of the eigenfunctions of the operators
$\CO^{(0)}_k=\frac{1}{(2\pi)^k}\Tr\,\varphi_0^k$ in the appropriate
polarization with the corresponding characters. The
obvious candidate  for the
replacement of \wavefunc\ in the Yang-Mills-Higgs theory is a set of
 wave-functions in the
$N$-particle sector of Nonlinear Schr\"{o}dinger  theory.
The basis in this space is defined in terms of the eigenfunctions of
the quantum Hamiltonian operator of Nonlinear Schr\"{o}dinger equation
and has an interpretation in terms of the representation theory of
  degenerate  (double) affine Hecke algebra.
 Before we consider this proposal in details let us discuss two limiting
cases of the
representation  \SDHPART, \SDHDIM\  that have  connections  with
representation
 theory of the classical Lie groups.

\subsec{$c\to \infty$}

In the limit $c\to \infty$ the $c$-dependent term in the action \ACTBOS\
 transforms into the delta-function of the
fields $\Phi$  and  the path  integral
reduces effectively to the path integral of the  two-dimensional
Yang-Mills theory discussed in Section 2. One should  expect that
the representation for the partition function \SDHPART\  to reproduce
 \PARTFUNC\ in this limit.  Indeed
in the limit $c\to \infty$ we have  $\mu(\l)\to 1$ and :
\eqn\LIMITO{
\lim_{c\to \infty} D_{\l}=\Delta_{\Fg}(\l)=\prod_{1\leq j<k\leq N}
(\l_j-\l_k),}
and $\l_i$ are the solutions of the limiting Bethe Ansatz  equations:
\eqn\BALIMITO{
(-1)^{N-1}e^{2\pi i\l_k}=1.}
Thus $\l_i=m_i+\frac{(N-1)}{2}$, $m_i\in \IZ$ and  one can
identify $m_i=\mu_i-i$ in the expressions \DIMGLN\ and \LIMITO.

\subsec{$c\to 0$}

In the opposite limit $c\to 0$ we  again have $\mu(\l)\to 1$ and :
\eqn\DIMINF{\lim_{c\to 0} c^{\frac{N(N-1)}{2}}\,D_{\l}=\Delta_{\Fg}(\l)^2=
\prod_{1\leq j<k \leq N}(\l_j-\l_k)^2,}
and $\l_i$ are  solutions of the limiting  Bethe Ansatz  equation:
\eqn\BAINF{
e^{2\pi i \l_k}=1.}
The interpretation of the limit is not so obvious because the localization
technique does not straightforwardly applicable for $c=0$.
However let us remind  that in the case of $c=0$ the theory \YMH\
is equivalent to the two-dimensional Yang-Mills theory with complex gauge
group $G^c$.  Thus one might expect that in the limit
$c\to 0$ on gets an answer with at least some
interpretation in terms of the
representation theory of the complexified group $G^c$.
In the Yang-Mills theory  for $G^c$ one expects to have a  sum
 (more exactly integral and the sum ) over the set of
unitary representations arising in the decomposition
of the regular representation of $G$  in  $L^2(G)$,
i.e.  over the  principal series of unitary
 representations. In order to compare this with the limit
$c\to 0$ let us first recall standard  facts
in  the representation theory of complex groups (see e.g.
\ref\VAR{V. S.~Varadarajan, {\it Harmonic Analysis on Real Reductive
    Groups}, Lecture Notes in Math.  576, Springer-Verlag, 1977.},
 \ref\JS{ D. P~Zelobenko, A. I.~Stern {\it
Predstavlenija grupp Li}, Moskow, Nauka, 1983.}).
For simplicity we discuss only the case $G=GL(N,\IC)$.
Principal series of unitary representations of $GL(N,\IC)$
can be studied  by inducing them from a Borel subgroup $B\subset
GL(N,\IC)$  using the character of $B$:
\eqn\BIND{
\chi(b)=\prod_{j=1}^N
|b_{jj}|^{i\rho_j-m_j}b_{jj}^{m_j}\,\,\,\,\,\,\,\,\,\,b\in
B,\,\,\,\rho_j\in
\IR,\,\,\,m_j\in \IZ.}
 For  complex groups all unitary representations are
 infinite-dimensional  and thus
the definitions of the character of the representation and
the the dimension of the representation deserves some care.
 The  character of the representation $\pi:G\to End(V)$ is defined as
follows.  Let $f(g)$ be a smooth  function with compact support. Then
define
the trace of $f$ as:
 \eqn\TRAGEN{
{\rm Tr}_V\,f\equiv{\rm Tr}(\int dg f(g) \pi(g)).}
Under some conditions \TRAGEN\ is well defined
(i.e. the operator \TRAGEN\ is of trace
class) and  one calls the generalized
function $\ch_V$ on $G^c$ a character if
\eqn\TRAGENone{
{\rm Tr}_V\,f=<{\rm ch}_{V},f>. }
It was shown by Harich-Chandra that thus defined generalized function
is an ordinary function and therefore ${\rm ch}_V$
 can be considered as a generalization of the
 characters of  finite dimensional representations.

The simplest example is a representation of $GL(N,\IC)$ obtained by
quantization of  the regular coadjoint orbit generalizing
two-sheet hyperboloid for $GL(2,\IC)$. The corresponding character is
given by:   \eqn\SLCHAR{
{\rm ch}_{\l}(e^x)=\frac{1}{|\Delta_G(e^x)|}
\sum_{w\in S_N}e^{2\pi i\sum_{j=1}^N\l_{w({j})}x_j}}
where $\l_j=m_j+i\rho_j$ and $S_N$ is a Weyl group of $GL(N,\IC)$.

In  the case of the finite-dimensional representations
the dimension of the representation is given by the value of the
corresponding character  at the unit element of the
group. However in the case of the infinite-dimensional representations
this relation can not be used to define the notion of dimension even
formally. In particular the value of the corresponding character
at the unit element can be infinite. For example
\SLCHAR\ tends to infinity when $x\to 0$ which is a
manifestation of the fact that the corresponding representation
is infinite-dimensional.

The correct definition of  the dimension $\CD_{\l}$
  of the principal series  unitary representations
is provided by the  decomposition:
\eqn\DELATDEC{
\delta_e^{(G)}(g)=\sum_{\l \in \widehat{G}} \,\CD_{\l}\, {\rm
ch}_{\l}(g),}
where $\delta_e^{(G)}(g)$ is a delta-function with the support at the unit
element $e\in G$  of the group, $\ch_{\l}(g)$ is a  character and
$\widehat{G}$ is a unitary dual to $G$ (i.e. the set of isomorphism
classes of the unitary representations entering the decomposition of the
regular representation).   The dimension $\CD_{\l}$ defined in such
way (known as a formal degree of the representation) coincides with the
ratio of the Plancherel measure arising
in the decomposition of the regular representation
 and the flat measure on the space of characters (see e.g. \VAR, \JS).
Explicitly we have  for $\CD_{\l}$:
\eqn\FORMDIMC{
\CD_{\l}=\prod_{1\leq j< k\leq N}|\l_i-\l_j|^2.}

Comparing \FORMDIMC\ with \DIMINF\ one infers that in the limit $c\to 0$
 one obtains the subset of the principal series of representations
corresponding to $\l_k=m_k\in \IZ$ (i.e. $\rho_k=0$).
It is reasonable to guess that in the limit $c\to 0$ the only information
that remains
is a class of functions and a particular class of representations
that arise in the spectral decomposition corresponds to this class of
functions\foot{This may  be compared with the Bogomolony limit of the
monopole equations where the only information on the potential that
survives in the limit is encoded in the asymptotic
behaviour of the solutions.}. This is natural because the  localization
demands a  compactification of the configuration space and $c\neq 0$ term
just
provides effectively this compactification. In the limit $c\to 0$
not all elements of $L^2(G)$ arise in the description of Hilbert space
of the theory  and we get a subset of the representations.

Thus the wave-functions of Yang-Mills-Higgs theory for $c\neq 0$
should interpolate between  characters of finite-dimensional
representations of $G$ and  characters
of a class of infinite-dimensional representations of
$G^c$. As we will demonstrate in the next Section
the wave-functions in the $N$-particle sector of Nonlinear Schr\"{o}dinger
theory provides exactly this interpolation.

\newsec{$N$-particle wave functions in  Nonlinear Schr\"{o}dinger  theory}

The appearance of a particular form of Bethe Ansatz equations \BA\
 strongly suggests the  relevance of  quantum integrable theories
in the description of  wave-functions in topological Yang-Mills-Higgs
theory. Precisely this form of Bethe Ansatz equations \BA\
 arises in the description of the $N$-particle wave functions
for the quantum Nonlinear Schr\"{o}dinger theory with the coupling
constant $c\neq 0$
\ref\LL{E. H. ~Lieb, W. ~Liniger, {\it Exact analysis of an interacting
Bose gas I. The general solution and the ground state}, Phys. Rev. (2),
 {\bf 130} (1963), 1605.},
\ref\BPF{F. A. ~Berezin, G. P. ~Pohil, V. M. ~Finkelberg, {Vestnik MGU}
{\bf
    1}, 1 (1964) 21.},
\ref\Y{C. N. ~Yang, {\it Some exact results for the many-body problems in
    one dimension with repulsive delta-function interaction},
Phys.Rev.Lett. {\bf 19} (1967), 1312.},
\ref\YY{C. N. ~Yang, C. P. ~Yang, {\it Thermodinamics of a one-dimensional
 system of bosons with repulsive delta-function interaction},
J. Math. Phys. {\bf 10} (1969), 1115.}.
In this  section we recall  the standard facts about the
construction of the these wave-functions
using the coordinate Bethe Ansatz. We also discuss the relation with
the representation theory of the degenerate (double) affine Hecke
algebras and the representation theory of the  Lie groups over complex and
$p$-adic numbers.
 For the application of the quantum inverse
scattering method to
 Nonlinear Schr\"{o}dinger theory see
\ref\KMF{P. P. ~Kulish, S. V. ~Manakov, L. D. ~Faddeev, {\it
Comparison of the exact quantum and quasiclassical  results for
the nonlinear Schr\"{o}dinger equation}, Theor. Math. Phys. {\bf 28}
(1976) 615.},
 \ref\FT{L. D. ~Faddeev, L. A. ~Takhtajan, {\it
Hamiltonian Methods in the Theory of Solitons}, Springer-Verlag,
(1980).}.
One can also recommend \ref\LDFONE{L. D.~Faddeev, {\it
How Algebraic Bethe Ansatz works for integrable model},
``Relativistic gravitation and gravitational radiation'', Proceedings of
Les Houches School of Physics (1995) p. 149,
[arXiv:hep-th/9605187].}
as  a quite readable introduction into the Bethe Ansatz machinery.

The Hamiltonian  of Nonlinear Schr\"{o}dinger theory  with a coupling
constant $c$  is given by:
\eqn\NLSH{
\CH_2=\int dx \,(\frac{1}{2}\frac{\pr \phi^*(x)}{\pr x}\frac{\pr \phi(x)}
{\pr x}+c(\phi^*(x)\phi(x))^2),}
with the following Poisson structure for bosonic fields:
\eqn\Poissonstr{
\{\phi^*(x),\phi(x')\}=\delta(x-x').}
The operator of the number of particles:
\eqn\PN{
\CH_0=\int dx \,\phi^*(x)\phi(x),}
commutes with the Hamiltonian $\CH_2$ and thus one can
 solve the eigenfunction problem
in the sub-sector for a given number of particles $\CH_0=N$.
We will consider  the  both the theory on
infinite interval ($x\in \IR$) and its periodic version $x\in S^1$.
The equation for eigenfunctions in the $N$-particle sector has the
following form:
\eqn\NPARTEIG{
(-\frac{1}{2}\sum_{i=1}^N\frac{\pr^2}{\pr
x_i^2}+c\sum_{1\leq i<j\leq N}\delta(x_i-x_j))\Phi_{\l}(x)=2\pi^2
(\sum_{i=1}^N\l_i^2)\Phi_{\l}(x)\,\,\,\,\,\,\,\,i=1,\cdots ,N.}
This equation is obviously symmetric with respect to the action of
symmetric group $S_N$  on the coordinates $x_i$. Thus the solutions are
classified according to the representations of $S_N$.
 Quantum integrability of the Nonlinear Schr\"{o}dinger
theory implies the existence of the complete set of the commuting
Hamiltonian operators. The corresponding eigenvalues
are given by the symmetric polynomials $p_k(\l)$.

Finite-particle sub-sectors of the Nonlinear Schr\"{o}dinger theory can be
described in terms of the representation theory of a particular kind
of Hecke algebra
\ref\MW{S. ~Murakami, M. ~Wadati {\it
Connection between Yangian symmetry and the quantum inverse scattering
method}, J. Phys. A: Math. Gen. {\bf 29} (1993) 7903.},
\ref\Hi{K. ~Hikami, {\it Notes on the structure of the -function
    interacting gas. Intertwining operator in the degenerate affine
    Hecke algebra}, J. Phys. A: Math. Gen  {\bf 31} (1998) No. 4.},
\ref\HO{G. J. ~Heckman, E. M. ~Opdam, {\it Yang's system of
particles and Hecke algebras}, Annals of Math. {\bf 145} (1997), 139.},
\ref\EOS{E. ~Emsiz, E. M. ~Opdam, J. V. ~Stokman,
{\it Periodic integrable systems with delta-potentials},
[arXiv:math.RT/0503034].}.
 Let $R=\{\a_1,\cdots,\a_{l}\}$ be a root system, $W$ - corresponding Weyl
group and
$P$ - a weight lattice. Degenerate affine Hecke algebra $\CH_{R,c}$
associated to $R$
is defined as an algebra with the basis $S_w$, $w\in W$
 and $\{D_{\l},\l\in P\}$  such that $S_w$ $w\in W$ generate
 subalgebra isomorphic to group algebra $\IC[W]$ and the elements
$D_{\l}$, $\l\in P$  generate the group algebra $\IC[P]$
 of the weight lattice $P$. In addition one has the relations:
\eqn\HA{
S_{s_i}D_{\l}-D_{s_i(\l)}\,S_{s_i}=c\,\frac{2(\l,\a_i)}{(\a_i,\a_i)},
\qquad\qquad \,\,\,\,\,i=1,\cdots
,n.}
Here $s_i$ are the generators of the Weyl algebra corresponding to the
reflection with respect to the simple roots $\a_i$. The center of
$\CH_{R,c}$
  is isomorphic to the algebra of $W$-invariant polynomial functions on
$R\otimes\IC$.
The degenerate  affine Hecke algebras were  introduced  by
Drinfeld \ref\Dr{V. G.~Drinfeld, {\it  Degenerate affine Hecke
algebras and Yangians}, Funct. Anal. Appl. {\bf 20} (1986) 58.} and
independently by Lusztig \ref\Lusztig{G.~Lusztig, {\it Affine Hecke
algebras and their
graded version}, J. AMS {\bf 2}, (1989) 3, 599.}.

Below we consider only the case of ${\frak gl}_N$ root system and thus
we have $W=S_N$. Let us introduce the following differential operators
(Dunkle operators
\ref\Dunkl{T. C.~Dunkl, {\it Differential-Difference operators
associated to reflection groups}, Trans. Amer. Math. Soc. {\bf 311},
 no.1 (1989), 167.}):
\eqn\Dunk{
\CD_i=-i\frac{\pr}{\pr x_i}+i\frac{c}{2}\sum_{j=i+1}^N
(\epsilon(x_i-x_j)+1)s_{ij}.}
Here $\epsilon(x)$ is a sign-function and $s_{ij}\in S_N$ is a
transposition
$(ij)$. These operators together with
the action of the symmetric group \Dunk\  provide a
representation of the degenerate affine algebra $\CH_{N,c}$ for
$\Fg={\frak gl}(N)$:
\eqn\REPDAH{
S_{s_i}\to s_i,\,\,\,\,D_i\to \CD_i,\,\,\,i=1,\cdots ,N.}
The image of the quadratic element of the center
is given by:
\eqn\QUADR{
\frac{1}{2}\sum_{i=1}^N\CD_i^2=
-\frac{1}{2}\sum_{i=1}^N\frac{\pr^2}{\pr x_i^2}+c\sum_{1\leq i<j\leq
N}\delta(x_i-x_j)}
and thus coincides with the restriction of the quantum Hamiltonian
on the $N$-particle sector of Nonlinear Schr\"{o}dinger theory on the
infinite interval.

 We are  interested in  $S_N$-invariant solutions of
 \NPARTEIG. They play the role of  spherical vectors (with respect to the
spherical
subalgebra $\IC[W]\in \CH_{N,c}$) in the representation theory
of degenerate affine Hecke algebra.

The eigenvalue problem \NPARTEIG\ allows the equivalent reformulation
as an eigenvalue problem in the domain $x_1\leq x_2 \leq \cdots \leq x_N$
for the differential operator:
\eqn\NPARTEIGFREE{
(-\frac{1}{2}\sum_{i=1}^N\frac{\pr^2}{\pr x_i^2})\Phi_{\l}(x)=2\pi^2
(\sum_{i=1}^N\l_i^2)\Phi_{\l}(x)\,\,\,\,\,\,\,\,i=1,\cdots ,N,}
 with the boundary conditions:
\eqn\BCN{
(\pr_{x_{i+1}}\Phi_{\l}(x)-\pr_{x_{i}}\Phi_{\l}(x))_{x_{i+1}-x_i=+0}
=4\pi c\Phi_{\l}(x)_{x_{i+1}-x_i=0}.}
The solution is given by:
\eqn\wfunc{
 \Phi^{(0)}_{\l}(x)=\sum_{w\in W}\,\,
\prod_{1\leq i<j\leq
N}\left(\frac{\l_{w(i)}-\l_{w(j)}+ic}{\l_{w(i)}-\l_{w(j)}}
\right)\exp(2\pi i\sum_k\l_{w(k)}x_k),}
 or equivalently:
\eqn\WFUNC{
 \Phi^{(0)}_{\l}(x)=\frac{1}{\Delta_{\Fg}(\l)}\,\sum_{w\in W}
 (-1)^{l(w)}\prod_{1\leq i<j\leq
N}(\l_{w(i)}-\l_{w(j)}+ic)\,\exp(2\pi i\sum_k\l_{w(k)}x_k).}
where $\Delta_{Fg}(\l)=\prod_{1\leq i<j\leq N}(\l_i-\l_j)$.
Note that the wave-function is explicitly symmetric
under the action of symmetric group $S_N$ on $\l=(\l_1,\cdots,\l_N)$.
This solution can be also  constructed using the representation
theory of degenerate affine Hecke algebra $\CH_{N,c}$ (see \MW, \Hi, \HO,
\EOS).

Given a solution of the equations \NPARTEIGFREE\ with boundary
conditions \BCN\  $S_N$-symmetric solutions of \NPARTEIG\ on $\IR^N$
  can be represented in the following form:
 \eqn\wfuncone{
 \Phi^{(0)}_{\l}(x)=\sum_{w\in W}
\left(\prod_{i<j}\left(\frac{\l_{w(i)}-\l_{w(j)}+ic\epsilon
(x_i-x_j)}{\l_{w(i)}-
\l_{w(j)}}
\right)\exp(2\pi i\sum_k\l_{w(k)}x_k)\right),}
where $\epsilon(x)$ is a sign-function.
This gives the full set of  solutions of  \NPARTEIG\
for $(\l_1\leq \cdots \leq \l_N)\in \IR^N$ satisfying the
orthogonality condition with respect to the natural pairing:
\eqn\PAIR{
<\Phi_{\l},\Phi_{\mu}>=\frac{1}{N!}\int dx_1\cdots dx_N\,
\overline{\Phi}_{\l}(x)
\Phi_{\mu}(x)=G(\l)\prod_{i=1}^N\delta(\l_i-\mu_i),}
where:
\eqn\NORMFAC{
G(\l)=\prod_{1\leq i<j\leq N}\frac{(\l_i-\l_j)^2+c^2}{(\l_i-\l_j)^2},}
Therefore the  normalized wave functions are given by:
\eqn\NORMWFUNC{
\Phi_{\l}(x)=\sum_{w\in W}(-1)^{l(w)}\,
\prod_{i<j}\left(\frac{\l_{w(i)}-\l_{w(j)}+ic\epsilon
(x_i-x_j)}{\l_{w(i)}-\l_{w(j)}-ic\epsilon(x_i-x_j)}\right)^{\frac{1}{2}}
\exp(2\pi i\sum_k\l_{w(k)}x_k).}
The eigenvalue problem for periodic $N$-particle
Hamiltonian of  Nonlinear Schr\"{o}dinger theory
 can be reformulated in the following way.
Consider the eigenfunction problem for the differential operator:
\eqn\NPARTEIGPeriod{
(-\frac{1}{2}\sum_{i=1}^N\frac{\pr^2}{\pr
x_i^2}+c\sum_{n\in \IZ}\,\sum_{1\leq i<j\leq
N}\delta(x_i-x_j+n))\Phi_{\l}(x)=2\pi^2
(\sum_{i=1}^N\l_i^2)\Phi_{\l}(x)\,\,\,\,\,\,\,\,i=1,\cdots ,N.}
The wave function of the periodic Nonlinear Schr\"{o}dinger equation
are the eigenfunction of \NPARTEIGPeriod\
satisfying the following invariance conditions:
\eqn\INVARAFFINE{
\Phi_{\l}(x_1,\cdots, x_j+1,\cdots, x_N)=\Phi_{\l}(x_1,\cdots, x_N),
\,\,\,\,\, j=1,\cdots ,N,}
$$
\Phi_{\l}(x_{w(1)},\cdots,
x_{w(N)})=\Phi_{\l}(x_1,\cdots,x_N),\,\,\,\,w\in S_N.$$
These are  the conditions of invariance under the action of the affine
Weyl group on the space of wave functions.

The solutions can be obtained imposing the
additional periodicity conditions on the
wave functions \NORMWFUNC. This  leads to the
following  set of the  Bethe Ansatz equations for
$(\lambda_1,\cdots ,\lambda_N)$:
\eqn\BAONE{
\CF_j(\l)\equiv e^{2\pi i \l_j}\prod_{k\neq
j}\frac{\l_k-\l_j-ic}{\l_k-\l_j+ic}=1,\,\,\,\,\,\,\,\,\,
\,\,j=1,\cdots ,N.}
The set of  solutions of these equations can be enumerated by
sets of integer numbers $(p_1 \geq \cdots  \geq p_N)$
- for each ordered set of these integers
there is exactly one solution to Bethe Ansatz equations \YY.
Let us remark that there is the following equivalent
representation for the periodic wave-functions:
\eqn\NORMWFUNCAFF{
\widetilde{\Phi}_{\l}(x)=\sum_{w\in W}(-1)^{l(w)}\,
\prod_{i<j}\left(\frac{\l_{w(i)}-\l_{w(j)}+ic}
{\l_{w(i)}-\l_{w(j)}-ic}\right)^{\frac{1}{2}+[x_i-x_j]}
\exp(2\pi i\sum_k\l_{w(k)}x_k),}
where $[x]$ is an integer part of $x$ defined by the conditions:
$[x]=0$ for $0\leq x<1$ and $[x+n]=[x]+n$. It easy to see that these
wave functions are periodic and descend to the wave functions \NORMWFUNC\
 if $\l=(\lambda_1,\cdots ,\lambda_N)$ satisfy \BAONE.
The  normalized wave functions in the periodic case are given by:
\eqn\PERIODNORM{
\Phi_{\l}^{norm}(x)=\left(\det\|\frac{\pr \log \CF_j(\l)}{\pr
    \l_k}\|\right)^{-1/2}\Phi_{\l}(x)=\mu(\l)^{-1/2}\Phi_{\l}(x).}
Note that the normalization factor is closely  related to the
factor \MEASURE\ arising in the representation of the partition
function of Yang-Mills-Higgs theory. Indeed the function $I(\l)$
introduced in \MEASURE\ is known in the theory of Nonlinear
Schr\"{o}dinger equations as Yang function \YY;
critical points of Yang function are in one to one correspondence with the
solutions
of Bethe Ansatz equations:
\eqn\YangF{ \alpha_j(\l)=\log
\CF_j(\l)=\frac{\pr I(\l)}{\pr \l_j}=n_j.}
 Below we will see that
this is not accidental. Finally note that the  periodic Nonlinear
Schr\"{o}dinger theory has an interpretation in terms of the
representation theory of the degenerate double affine Hecke
algebras introduced by Cherednik \ref\ICH{I. ~Cherednik, {\it A
unification of Knizhnik-Zamolodchikov and Dunkle operators via
affine Hecke algebras}, Invent. Math. J. {\bf 106} (1991), 411.}.
For the details in this regard see \EOS.

\subsec{$c\to \infty$: Representation theory of compact Lie groups}

The limit $c\to \infty$ corresponds to the case of impenetrable
bosons and the correlation functions are naturally
 represented in terms of free fermions. Indeed in
the limit $c\rightarrow \infty$  Bethe Ansatz  equations are reduced
to the condition:
\eqn\ANTIPER{
\l_j=\frac{N-1}{2}+m_i,\,\,\,\,\,\,\,\,\,m_i\in \IZ,}
and the wave-function is given by the
 wave-function of free fermions (up to a simple sign factor):
\eqn\COLIMIT{
\Phi^{c=\infty}_{\l}(x)=\frac{|\Delta(e^x)|}{\Delta(e^x)}\,
\det\|e^{i\l_k x_j}\||=\frac{|\Delta(e^x)|}{\Delta(e^x)}\,\sum_{w\in
  W}(-1)^{l(w)}e^{2\pi i\sum_{k=1}^
N\l_{w(k)}x_k}.}
Note that we have a simple relation with the characters
of the finite-dimensional representations of $U(N)$:
\eqn\CHNLS{
\ch_{\l}(x)= \frac{1}{|\Delta(e^x)|}\Phi^{c=\infty}_{\l}(x).}
Thus in the limit $c\to \infty$ the wave function $\Phi_{\l}(x)$
in Nonlinear Schr\"{o}dinger theory  can be considered as  a wave-function
in
two-dimensional Yang-Mills  theory renormalized according to
$\Phi_i(x)=|\Delta_G(e^x)|\Psi_i(x)$ (see Section 2).

\subsec{$c\to 0$: Representation theory of complex Lie groups}

In the limit $c\rightarrow 0$  Bethe Ansatz  equations are reduced
to the condition:
\eqn\PER{
\l_i=m_i,\,\,\,\,\,\,\,\,\,m_i\in \IZ,}
and the wave-functions are  given by:
 \eqn\CINFLIMIT{
\Phi^{c=0}_{\l}(x)=\sum_{w\in W}(-1)^{l(w)}e^{2\pi i\sum_{k=1}^
N\l_{w(k)}x_k}.}
Note that  wave functions \CINFLIMIT\ are normalized with respect
to the standard scalar product. Now we have a  simple relation with the
characters
\SLCHAR\ of the infinite-dimensional representations of $GL(N,\IC)$:
\eqn\INFCR{
\ch_{\l}(x)=\frac{1}{|\Delta(e^x)|}\Phi^{c=0}_{\l}(x).}

\subsec{$c\neq 0, \infty$: Representation theory of $p$-adic Lie groups}

Wave functions of Nonlinear Schr\"{o}dinger theory  for $c\neq 0,
\infty$  has  also a connection  with  the representation theory of Lie
groups due to the relation between  the representation theory of the
degenerate affine Hecke algebras $\CH_{N,c}$ and  the representation
theory of $GL(N,\IQ_p)$ where $\IQ_p$ is a field of $p$-adic numbers.
More specifically the wave functions of Nonlinear Schr\"{o}dinger theory
can be obtained as a limit of Hall-Littlewood polynomials that can be
considered as  generalized  zonal spherical functions
for $GL(N,\IQ_p)$. The limit is a kind of $p\to 1$ limit\foot{For another
example
where the same limit $p \rightarrow 1$ is relevant
in  string theory  see \ref\gs{Anton A.  Gerasimov and
Samson L. Shatashvili, {\it On Exact tachyon
Potential in Open String Field Theory}, JHEP 0010:034,2000,
hep-th/0009103.}.}. For the detailed discussion  of
$p$-adic zonal spherical functions see
Macdonald \ref\MACD{I.~Macdonald, {\it Harmonic   Analysis on
 Semi-simple Groups},  Actes, Cong\`{e}s Intern. Math.,
  {\bf 2} (1970), 331.}
 and for the relation with Nonlinear Schr\"{o}dinger theory
through representation theory of degenerate Hecke algebras
see \HO.

Let us start with the definition of Hall-Littlewood polynomials
(see e.g. \ref\MCD{I. G.~Macdonald, {\it Symmetric
    Functions and Hall Polynomials}, Oxford University Press, London,
  1979.}). Let $\{\Lambda_i\}$, $i=1,\cdots ,N$  be a set of formal
variables and $\mu=(\mu_1,\cdots ,\mu_N)$ be a partition of length
$N$. Then Hall-Littlewood polynomial depending on additional
formal variable $t$ is defined as:
\eqn\HL{
P_{\mu}(\Lambda_1,\cdots ,\Lambda_N|t)=\frac{1}{v_{\mu}(t)}
\sum_{w\in S_N}\,w\left(\Lambda_1^{\mu_1}\cdots \Lambda_N^{\mu_N}
\prod_{i<j}\frac{\Lambda_i-\Lambda_jt}{\Lambda_i-\Lambda_j}\right)=}
 $$
=\frac{1}{v_{\mu}(t)\,\Delta(\Lambda)}\,\sum_{w\in S_N}\,(-1)^{l(w)} \,w\,
(\Lambda_1^{X_1}\cdots \Lambda_N^{X_N}
\prod_{i<j}(\Lambda_i-\Lambda_jt)),
$$
where for the partition $\mu=(1^{m_1},2^{m_2},\cdots ,r^{m_r},\cdots )$:
  \eqn\NEWFUNC{
v_{\mu}=\prod_{j=1}^{N}\,\prod_{i=1}^{m_j}\frac{1-t^i}{1-t},\qquad \qquad
\Delta(\Lambda)=\prod_{i<j}(\Lambda_i-\Lambda_j).}
Hall-Littlewood polynomials enter the explicit formulas
for the zonal spherical functions for $p$-adic Lie groups.
The zonal spherical functions are defined as follows.
Given a $p$-adic Lie group $G$ and its  maximal compact subgroup
$K\subset G$ zonal spherical function $\omega(g)$ on $G$ is a  continuous
complex-valued functions satisfying the following conditions: (1)
the function is invariant with respect to the left and right action of
the compact subgroup $\omega(kgk')=\omega(g)$, $k,k'\in K$, (2)
normalization condition $\omega(1)=1$, (3) the function
 is eigenfunction for the convolution with any function with compact
support on $G$ satisfying (1). The spherical functions for
$G=GL(N,\IQ_p)$,
$K=GL(N,\IZ_p)$ (here  $\IZ_p$ is a ring of $p$-adic integers)
has the following representation in terms Hall-Littlewood
polynomials. Note that the set of the representatives of the double-coset
$K\backslash G/K$ can be identified with the elements of the form
$(p^{\mu_1},\cdots, p^{\mu_N})\in K\backslash G/K$ where
$\mu=(\mu_1\geq \mu_2\geq \cdots \geq \mu_N)$ is a partition.
Then for the zonal spherical functions we have:
\eqn\SPHERF{
\omega_{s}(p^{\mu_1},\cdots,p^{\mu_N})=p^{-\sum_{i=1}^N
(n-i)\mu_i}\frac{v_{\mu}(p^{-1})}
{v_{N}(p^{-1})}P_{\mu}(p^{-s_1},\cdots,p^{-s_N}|p^{-1}),}
where $s=(s_1,\cdots ,s_N)\in \IZ^N$ and $$
v_{N}(t)=\prod_{i=1}^{N}\frac{1-t^i}{1-t}.$$
Note  that the ``spectral'' indexes in the spherical function and
Hall-Littelwood  polynomial are interchanged.
There is a generalization of the notion of the spherical function
which is similar to the multivariable  hypergeometric functions
for general root systems introduced by Heckman and Opdam (see e.g.
\ref\GHHS{G.J.~Heckman and H.~Schlichtkrull,{\it  Harmonic Analysis
Functions on Symmetric Spaces}, Academic Press, 1994.}). For the case
of the $p$-adic Lie groups the generalized spherical function depends
on the additional integer  parameter $k\in \IZ$ and is given by:
\eqn\SPHERFK{
\omega^{(k)}_{s}(p^{\mu_1},\cdots,p^{\mu_N})=p^{-\sum_{i=1}^N
(n-i)\mu_i}\frac{v_{\mu}(p^{-k})}
{v_{N}(p^{-k})}P_{\mu}(p^{-s_1},\cdots,p^{-s_N}|p^{-k}).}

Now one can see how  in a particular limit
the wave functions of $N$-particle sector of Nonlinear Schr\'{o}dinger
theory
on $\IR$  obtained using the coordinate Bethe Ansatz arise.
Taking   $\mu_i=\epsilon^{-1}   x_i$,
$\Lambda_i=\exp(2\pi \epsilon \l_i)$  and $t=e^{2\pi i c\epsilon}$ while
$\epsilon\to 0$
(we use analytical continuation over parameters here)
we have for \HL:
\eqn\REMAC{
 \frac{v_{\mu}(t)}
{v_{N}(t)}P_{\mu}(\Lambda_1,\cdots,\Lambda_N;t)\to
 \frac{1}{(c\epsilon)^N} \frac{1}{N!}\sum_{w\in S_N}\,w\left(e^{2\pi
i\sum_{k=1}^N x_k\l_k}
\prod_{k<j}\frac{\l_k-\l_j+ic}{\l_k-\l_j}\right)}
This expression coincides with the restriction of the Bethe wave
function on the subspace  $x_1<x_2<\cdots <x_N$.
Thus taking into account \SPHERFK\ one can conclude that the
generalized zonal spherical functions in the formal limit
$\epsilon\to 0$ while $p=e^{2\pi i \epsilon}$,
$\mu_i=\frac{1}{\epsilon}x_i$ are given by the wave functions for the
$N$-particle sector of Nonlinear Schr\"{o}dinger equation with $c=k$.

It known that Hall-Littlewood polynomials are a special case of the
Macdonald polynomials and one can expect that Hall-Littlewood
polynomials before degeneration and more general Macdonald polynomials
should be related with the quantum integrable/topological
field theories these  along the line discussed in this section.
Below in Section 8 we propose the generalization
of the topological Yang-Mills-Higgs theory that provides
a realization of these more general polynomials.

\newsec{Wave-function in topological Yang-Mills-Higgs theory}

In this section we provide the evidences for the identification of
a bases of the wave functions of topological Yang-Mills-Higgs
theory for $G=U(N)$  (given by a
 path integral on a disk with the insertion of observables in the
center)  with the eigenfunctions of the  $N$-particle
Hamiltonian operator  of Nonlinear Schr\"{o}dinger theory.
  First by counting the observables of
the theory we show that the phase space of the Yang-Mills-Higgs
theory can be considered as a deformation of the phase space of
Yang-Mills theory.
This implies that the bases of wave functions
in  Yang-Mills-Higgs theory can be obtained by a deformation
of the bases of wave functions in Yang-Mills theory.
 Next, using the explicit representation
of the partition function on the two-dimensional torus we derive the
transformation properties of the wave functions under large gauge
transformations. They are in agreement with the known explicit
transformation properties of  wave function in Nonlinear
Schr\"{o}dinger theory. Finally, we compute the cylinder path
integral (Green function) and torus partition function in Nonlinear
Schr\"{o}dinger theory (with all higher Hamiltonians) and show that
latter coincides with torus partition function in Yang-Mills-Higgs
theory (with arbitrary observables turned on). Taking into account
that the constructed wave functions in the gauge theory are the
eigenfunctions  of the full set of the Hamiltonian operators these
considerations
presumably uniquely fix the set of wave functions.

\subsec{Local $Q$-cohomology}

We start with a description of the Hilbert space of the
Yang-Mills-Higgs theory using the operator-state correspondence.
In the simplest form the operator-state correspondence is as
  follows. Each operator, by acting on the  vacuum state,
 defines a sate in the Hilbert space. In turn for each state
there is an operator, creating the state from the vacuum state. Moreover,
for the maximal commutative subalgebra of the operators this
correspondence should be  one to one.
For example, the space of local gauge-invariant $Q$-cohomology classes
in topological $U(N)$ Yang-Mills theory is spent, linearly, by the
operators:
 \eqn\OPYM{
\CO^{(0)}_k=\frac{1}{(2\pi i)^k}\Tr\,\varphi^k,}
and  thus this space coincides with the space of  ${\rm Ad}_G$-invariant
regular functions on the Lie algebra ${\frak u}_N$.
 This is in accordance with the description of
the Hilbert space of the theory  given in Section 2.

We would like to apply the same reasoning to  the topological
Yang-Mills-Higgs theory. To get   economical description of the Hilbert
space of
the theory one should find  a maximal (Poisson) commutative
subalgebra of  local $Q$-cohomology classes
(where $Q$ given by \BRST\ acts on the space of functions
invariant under the symmetries generated by \UONEACT, \GAUGEYM).
Obviously  operators \OPYM\ provide non-trivial cohomology
classes. One can show that these operators provide a maximal
 commutative subalgebra for $c\neq 0$ and therefore
 the reduced phase space in Yang-Mill-Higgs system
 can be identified  with a  phase space of  pure Yang-Mills theory.
Thus the  Hilbert space of Yang-Mills-Higgs theory ($c\neq 0$)
can be naturally identified with the Hilbert space of
Yang-Mills theory (identified with $c\to \infty$).
The fact that the Hilbert space of Yang-Mills-Higgs
 theory is the same for all  $c\neq 0$ implies that the bases of
wave-functions for $c\neq 0$ should be
a deformation of the bases for  $c=\infty$.

One should  stress that this reasoning is not applicable to the case $c=0$
\foot{We distinguish the case  corresponding to Yang-Mills
theory for a complex group ($c=0$)  and $c\to 0$ in the Yang-Mills-Higgs
theory. In the latter case, as we explained before, the Hilbert space is
the
same as for Yang-Mills-Higgs theory with $c\neq 0$ and
the corresponding bases of wave functions is given by the formal
characters \SLCHAR\ for a subset of the unitary representations
 of the complex group.}.
The local cohomology for $c=0$ contains additional operators.
 For example, the following operators provide non-trivial cohomology
classes for arbitrary $t\in \IC$ and $c=0$:
\eqn\OPYNC{
\CO^{(0)}_k(t)=\frac{1}{(2\pi i)^k}\Tr(
\varphi_0+t\varphi_+-\frac{t}{2}\chi_+^2)^k.}
This is a manifestation of the fact that $c=0$ theory is a Yang-Mills
  theory for the complexified group $G^c$ and thus its phase space is
  given by $\CM_{\IC}=T^*H^c/W$.

The identification of the Hilbert spaces of Yang-Mills-Higgs theory
and of pure Yang-Mills theory  supports the idea to
use  wave functions in Nonlinear Schr\"{o}dinger
theory discussed in the previous section as a bases in the Hilbert
space of Yang-Mills-Higgs theory.  Below we provide
 further evidences for this identification.

\subsec{Gauge transformations  of wave function}

The discreteness of the  spectrum of  $N$-particle Hamiltonian
operator in periodic Nonlinear Schr\"{o}dinger theory
arises due to periodicity condition on wave functions.
Thus the eigenfunctions in the periodic  case are given by a subset of
eigenfunctions  on $\IR^N$ descending  to $S_N$-invariant
functions on $(S^1)^N$.
The eigenfunction \NORMWFUNC\  of the Hamiltonian operator on
$\IR^N$ represented as sum over elements of symmetric group
of simple wave  functions. For generic eigenvalues each  term of
the sum is multiplied by some function under the shift $x_i\to
x_i+n_i$, $n_i\in \IZ$ of the coordinates. Below we will show how these
multiplicative factors arising in Nonlinear Schr\"{o}dinger theory
can be derived in Yang-Mills-Higgs gauge theory.

We start with a simple  case of  Yang-Mills theory.
The partition function of $U(N)$ Yang-Mills theory on the torus
$\Sigma_1$ is given by:
\eqn\YMMMsumone{
Z_{YM}(\Sigma_1)=\int_{\IR^N/S_N} d^N\l\,\sum_{(n_1,\cdots ,n_N)\in \IZ^N}
e^{2\pi  i \sum_{m=1}^N\l_m n_m}
\,\,e^{-\sum_{k=1}^{\infty}\,t_k\,\,p_{k}(\l)}=}
$$=\sum_{(m_1,\cdots ,m_N)\in P_{++}}
\,\,e^{-\sum_{k=1}^{\infty}\,t_k\,\,p_{k}(m+\rho)}$$
where $P_{++}$ is a set of the dominant weights of $U(N)$.
The sum over $(n_1,\cdots ,n_N)\in \IZ^N$ has a meaning
of the sum over topological classes of
$U(1)^N$-principle bundles on the torus $\Sigma_1$.
 It results in the replacement  of the
integration over $\l$ by a sum over a discrete subset.
This should be compared with the  partition function of dimensionally
 reduced $U(N)$  Yang-Mills theory on $S^1$:
\eqn\YMREDPARTFUNC{
Z_{QM}(S^1)=\int_{\IR^N/S_N} d^N\l
\,\,e^{-\sum_{k=1}^{\infty}\,t_k\,\,p_{k}(\l)}.}
Contrary to the two-dimensional Yang-Mills theory
in the last case we do not have any additional restriction on the
spectrum $(\l_1,\cdots ,\l_N)\in \IR^N/S_N$.
The appearance of the additional sum in \SDHsumone\
can be traced back to the difference between
the Hilbert spaces of dimensionally reduced and non-reduced theories.

 The mechanism of the spectrum restriction via the sum over the
topological sectors
 can be explained in terms of the structure of the Hilbert space of the
theory as follows.  In the Hamiltonian formalism the partition function
on a torus is given by the  trace
of the evolution operator over the Hilbert space of the theory.
Let us consider first the dimensionally reduced $U(N)$ Yang-Mills theory.
The  phase space of the theory is  $T^*\IR^N/S_N$ where we divide over
Weyl group $W=S_N$. To construct the Hilbert
space we quantize the phase space using  the following   polarization.
Consider Lagrangian projection  $\pi: T^*\IR^N\to\IR^N$
supplied with a section. We chose the  coordinates
on the base  as position variables and
the  coordinates on the fibers  as the corresponding  momenta.
Thus  the Hilbert space in this polarization is realized as a space of
$S_N$ (skew)-invariant functions on the base $\IR^N$ of the projection.

Now  consider  two-dimensional Yang-Mills
theory. For the phase space we have  $T^*H/S_N$ where
$H$ is Cartan subgroup. We use $\,$ similar
$\,$ polarization $\,$ associated $\,$ with $\,$
the projection $\,\,\,\,\,$ $\pi: T^*H\to H$.
 Thus the wave functions are $S_N$ invariant functions on
a  torus $H$ or equivalently  the functions on $\IR^N$
invariant under action of the semidirect product of the lattice
$P_0=\pi_1(H)$ and Weyl $W=S_N$ group (i.e. under the action of the
affine Weyl group $W^{aff}$).
The lattice $P_0$ can be interpreted as a lattice of the
 $\IR^N$-valued constant connections on $S^1$ which are  gauge equivalent
to the
zero connection. The corresponding gauge transformations
act on the  wave functions   by the shifts
$x_j\to x_j+ n_j$, $n_j \in \IZ$  of the argument
of the wave functions in the chosen polarization and the wave
functions in two-dimensional Yang-Mills theory can be obtained by the
averaging over this gauge transformations and global gauge
transformations by the nontrivial elements of the normalizer of Cartan
torus $W=N(H)/H$.

It  is possible to relate the averaging
over the topologically non-trivial transformations with the
sum over topological classes of $H$ bundle on the tours.
Note that the maps of $S^1$ to the gauge group $H$
are topologically classified by $\pi_1(H)=\IZ^N$.
Consider  a connection  $A=(A_1,\cdots, A_N)$
 on a $H$ bundle  over  a cylinder  $L$,
$\pr L=S^1_+\cup S^1_-$  such that the holonomies along the boundaries
$S^1_+$ and $S^1_-$ are in the different topological classes
$[(m_1,\cdots,m_N)]\in \pi_1(H)$ and $[(m_1+n_1,\cdots, m_N+n_N)]\in
\pi_1(H)$.
Gluing  boundaries  of the cylinder $L$ we obtain  a torus supplied
with a  connection $\nabla_A$ such that the  first Chern classes of
the bundles corresponding to each $U(1)$-factor  are given by
 $c_1(\nabla_{A_i})=\frac{1}{2\pi i}\int_LF(A_j)=n_j$, $j=1,\cdots, N$.
Thus we see that the sum over the topologically non-trivial gauge
transformations on $S^1$ can be translated into the sum over
topological classes of the $H$-bundles on the torus.

Let us re-derive  the partition function of Yang-Mills theory on the
torus \YMMMsumone\ using the averaging procedure. We start with the
dimensionally reduced theory. Let us chose a bases in the Hilbert
space of the dimensionally reduced Yang-Mills theory given by the
$S_N$ skew-invariant eigenfunctions of the quadratic operator
$H^{(0)}_2=\tr \varphi^2$. In the polarization discussed above we
have:
 \eqn\HAMONE{
H^{(0)}_2\,\psi_{\l}(x)=-\frac{1}{2}\left(\sum_{j=1}^N\frac{\pr^2}{\pr
  x_j^2}\right)\,\psi_{\l}(x)=2\pi^2\sum_{j=1}^N\l_i^2\,\psi_{\l}(x),}
where $(x_1,\cdots,x_N)\in \IR^N$.
The set of  normalized skew-invariant eigenfunctions is given by:
\eqn\NORMEIGONEDIM{
\psi_{\l}(x)=\sum_{w\in S_N}\,(-1)^{l(w)}\,
\exp(2\pi i \sum_{j=1}^N\l_{w(j)} x_j),\qquad \,\,\,\,\,\,\,\,\,\,\,
(\l_1,\cdots ,\l_N)\in \IR^N/W.}
\eqn\NORMILAIZATION{
\frac{1}{N!}\int_{\IR^N}\,\,d^Nx\,\,
\overline{\psi_{\l}}(x)\,\psi_{\l'}(x)=(2\pi)^N
 \sum_{w\in S_N}\,(-1)^{l(w)}\prod_{j=1}^N\delta(\l_{w(j)}-\l'_j)
=\delta^{(S_N)}(\l-\l').}
The integral kernel of the identity operator acting
on the skew-symmetric functions can be represented
(due to translation invaraince
it is the function of difference $x-x'$) as:
\eqn\KERNEL{
K_0(x,x')=K_0(x-x')=\delta^{(S_N)}(x-x')=\int_{\IR^N/S_N}\, d^N\l\,\,
 \overline{\psi_{\l}}(x)\,\psi_{\l}(x').}
The partition function of the dimensionally reduced Yang-Mills theory
on $S^1$ is given by the trace of a evolution operator and
can be written explicitly as:
\eqn\TRACEOP{
Z_{QM}(S^1)=Tr\, e^{-t_2 H_2(\hat{p},\hat{q})}=
\int_{(\IR^N\times \IR^N)/S_N}\,d^N x\, d^N\l\,\,\,\,
 \overline{\psi_{\l}}(x)\,e^{-t_2H^{(0)}_2(i\pr_x,x)}
\,\psi_{\l}(x)=}
$$
=\int_{(\IR^N\times \IR^N)/S_N}\,d^N x\, d^N\l\,\,\,\,
 e^{-t_2 p_2(\l)}.$$
The Green function of the theory is:
\eqn\gzero{G_0(x,x')=\int_{\IR^N/S_N} d^N\l\,\,\,\,
 \overline{\psi_{\l}}(x')\,e^{-t_2H^{(0)}_2(i\pr_x,x)}
\,\psi_{\l}(x)=}
$$
=\int_{\IR^N/S_N} d^N\l\,\,\,\, \psi_{\lambda}(x')
 e^{-t_2 p_2(\l)}\psi_{\lambda}(x).$$
Up to the infinite factor given by the integral over
$x=(x_1,\cdots,x_N)\in \IR^N$ the integral in \TRACEOP\ coincides with the
expression \YMREDPARTFUNC\ for the partition function for $t_{i\neq 2}=0$.

Now consider two-dimensional Yang-Mills theory.
In this case we have the periodic eigenvalue problem for \HAMONE. Then for
the
normalized eigenfunctions of $H_2$ we have:
\eqn\NORMEIGONEDIMper{
\psi_{n}(x)=\sum_{w\in S_N}\,(-1)^{l(w)}\,
\exp(2\pi i \sum_{j=1}^N (n_{w(j)}+\rho_{w(j)}) x_j),
\qquad \,\,\,\,\,\,\,\,\,\,\,(n_1,\cdots,n_N) \in P_{++},}
\eqn\NORMILAIZATIONper{
\frac{1}{N!}\int_{(S^1)^N}\,\,d^Nx\,\,
 \overline{\psi_{n}}(x)\,\psi_{n'}(x)=\sum_{w\in S_N} \,(-1)^{l(w)}
\prod_{j=1}^N \delta_{n_{w(j)},n'_j}=\delta^{S_N}_{n,n'}.}
Here $\rho=(\rho_1,\cdots ,\rho_N)$ is a half-sum of the positive
roots of  $\frak{u}_N$.
 The integral kernel of the identity  operator can be represented as:
\eqn\KERNELper{
K(x,x')=K(x-x')=\delta_{(S_N)}(x-x')=
\sum_{n\in P_{++}}\,\, \overline{\psi_{n}}(x)\,\psi_{n}(x').}
The partition function of the Yang-Mills theory
on a torus $\Sigma_1$ is given by the trace of a evolution operator and:
\eqn\TRACEOP{
Z_{YM}(\Sigma_1)=
Tr\, e^{-t_2 H_2(\hat{p},\hat{q})}=\sum_{n\in
P_{++}}\,\,\int_{(S^1)^N}\,d^Nx \,\,
 \overline{\psi_{n}}(x)\,e^{-t_2 H^{(0)}_2(i\pr_x,x)}\,\psi_{n}(x).}
The kernel for the periodic case can be obviously represented as a matrix
element of the projection operator as follows:
\eqn\KERNELperiod{
K(x,x')=\int_{\IR^N/S_N}\, d^N\l\,\,
 \overline{\psi_{\l}}(x)\,\,\,P(\l)\,\,\psi_{\l}(x'),}
where the  wave-functions $\psi_{\lambda}(x)$ are given by \NORMEIGONEDIM\
and:
\eqn\PROJECTOR{
P(\l)=\sum_{m\in \IZ^N}\,\,\prod_{j=1}^N \delta(\l_j-m_j)=
\sum_{k\in \IZ^N}\,\,e^{2\pi i \sum_{j=1}^N\l_j k_j}.}
Equivalently we have:
\eqn\KERNELperiodsum{
K(x,x')=\sum_{k\in \IZ^N}
\int_{\IR^N/S_N}\, d^N\l\,\,
 \overline{\psi_{\l}}(x)\,\,\,e^{2\pi i\sum_{j=1}^N \l_j
   k_j}\,\,\psi_{\l}(x')=}
$$=\sum_{k\in \IZ^N}
\int_{\IR^N/S_N}\, d^N\l\,\,
 \overline{\psi_{\l}}(x)\,\,\, \psi_{\l}(x'+k).$$
We conclude that the Green function $G(x,x')$
(the path integral on the cylinder with insertion
of $\exp{(-t_2 H^{(0)}_2)}$)  is represented as:
\eqn\cyl{G_{YM}(x,x')= \sum_{n\in \IZ^N}\,\,\int_{\l\in
\IR^N/S_N}d^N\l\,\,
 \overline{\psi_{\l}}(x)\, \,e^{2\pi i\sum_{j=1}^N \l_j n_j} e^{-t_2
p_2(\l)}\,\,
\psi_{\l}(x')}
or equivalently as:
\eqn\cylone{G_{YM}(x,x')= \sum_{k\in \IZ^N}\,\,\int_{\l\in
\IR^N/S_N}d^N\l\,\,
 \overline{\psi_{\l}}(x) e^{-t_2 p_2(\l)}\,\,
\psi_{\l}(x'+k).}
Let us  note that the identities  in \KERNELperiodsum, \cylone\ are
 based on the following transformation  property of the  complete
set of skew-symmetric normalized wave-functions on $\IR^N$:
\eqn\propert{\psi_{\lambda}(x+k)=
\sum_{w\in S_N} (-1)^{l(w)}\,
e^{2\pi i \sum_{j=1}^N\l_{w(j)} k_j}
 e^{2\pi i \sum_{j=1}^N\l_{w(j)} x_j}.}
Thus each elementary term in the averaging over $S_N$ is
multiplied on the simple exponent factor entering the description
of the projector \PROJECTOR. Let us also note that the shift
transformations in \propert\ can be interpreted as large gauge
transformations in Yang-Mills theory discussed above.

The representation \cylone\ can be written in the following form:
\eqn\green{G_{YM}(x,x')=\sum_{k\in \IZ^N}G_0(x,x'+k).}
If  we set the coupling $t_2$ to zero, $t_2=0$, we recover the
formula \KERNELperiodsum\ for
$K(x,x')$:
\eqn\for{K(x,x')=\sum_{k\in \IZ^N}K_0(x,x'+k).}
For the partition function of Yang-Mills theory on a torus we get
(after setting $x=x'$ above and integrating over $x$):
\eqn\TRACEOPPP{
Z_{YM}(\Sigma_1)=\sum_{n\in \IZ^N}\,\,\int_{\l\in \IR^N/S_N}d^N\l\,
 \int_{(S^1)^N}\,d^Nx \,\,
 \overline{\psi_{\l}}(x)\, \,e^{2\pi i\sum_{j=1}^N \l_j n_j} e^{-t_2
p_2(\l)}\,\,
\psi_{\l}(x)=}
$$
=\sum_{m\in P_{++}}\,\, e^{-t_2 p_2(m+\rho)}.
$$
and this coincides with the representation \YMMMsumone. Note that
obvious relations between \KERNELperiodsum,  \green, \for\  and the
averaging
over the topologically non-trivial gauge transformations discussed
above.

Let us remark that the averaging procedure represented by \green,
\for\ is a standard tool in construction of Green functions on
non-simply connected spaces. At the first step one computes
the Green function on the universal
covering space and then  averages with respect to the action of
 the action of $\pi_1$ of the underlying non-simply connected space.
For example this procedure  has been used in similar problem  of
the quantization of  the coadjoint orbits of compact Lie groups
 in \ref\afs{A.~Yu.~Alekseev, L.~ D.~Faddeev and
S.~L.~Shatashvili, {\it  Quantization of symplectic
orbits of compact Lie groups by means of the functional integral,}
J. Geom. Phys., 5:391-406, 1988.}. We will apply this procedure to the
Yang-Mills-Higgs theory in a fashion
described above for 2d Yang-Mills theory.

Now we are finally ready to consider the case of Yang-Mills-Higgs
theory.
As it was conjectured above one can chose as a bases of the wave-functions
the bases of eigenfunctions of the set of the Hamiltonian operators
in $N$-particle subsector of Nonlinear Schr\"{o}dinger theory.
Below we construct the Green function and partition function
in Nonlinear Schr\"{o}dinger theory
and demonstrate that identifying the Hamiltonian operator with
quadratic observable $\CO^{(0)}_2=\frac{1}{(2\pi)^2}{\rm Tr} \varphi_0^2$
in Yang-Mills-Higgs theory we reproduce the partition function of
Yang-Mills-Higgs theory on a torus.

Let us start with the construction of the kernel of the unit
operator in the bases of the $N$-particle eigenfunctions
of the Nonlinear Schr\"{o}dinger theory.
The representation for the kernel \KERNELperiodsum\
can be straightforwardly generalized to this case:
\eqn\KERNELperiodsum{
\widetilde{K}(x,x')=\sum_{(\l_1,\cdots \l_N)\in \CR_N}\
 \overline{\Phi_{\l}^{norm}}(x)
 \Phi^{norm}_{\l}(x')=\int_{\IR^N/S_N}\,\,d^N
\l\,\,\,\,\,\overline{\Phi_{\l}}(x)
\, P(\l) \,\Phi_{\l}(x'),}
 where $\Phi_{\l}(x)$ are normalized skew-invariant eigenfunctions on
$\IR^N$ given by \NORMWFUNC, $\Phi^{norm}_{\l}(x)$ are
normalized periodic eigenfunctions
 given by \PERIODNORM\ and the sum goes over the set $\CR_N$ of the
solutions of Baxter Ansatz equations.  The projector here is given by:
\eqn\PROJKERNELPERIOD{
P(\l)=\mu(\l) \,\sum_{m\in \IZ^N}\,\prod_{j=1}^N\,\,\delta(\a_j(\l)- m_j)=
\sum_{(\l^*_1,\cdots,\l^*_N)\in \CR_N}\prod_j\delta(\l_j-\l_j^*)}
where  $\a_j(\l)$ are defined as follows  (compare with \yang):
\eqn\YYANG{\alpha_j(\lambda)=\l_j+\frac{1}{2\pi i}
\sum_{k\neq j}\log \left(\frac{\l_k-\l_j-ic}{\l_k-\l_j+ic}\right).}
Then we have:
$$
\widetilde{K}(x,x')=\sum_{n\in \IZ^N} \int_{\IR^N/S_N}\,
d^N\l\,\,\mu(\l)\,\,
 \overline{\Phi}_{\l}(x)\,\,\,
e^{2\pi i\sum_{m=1}^N \l_m n_m}\,\prod_{l\neq j}
\left(\frac{\l_l-\l_j-ic}{\l_l-\l_j+ic}\right)^{n_j}
\,\,\Phi_{\l}(x')=$$
\eqn\PROJKERNELPERIODD{
=\sum_{k\in \IZ^N}
\int_{\IR^N/S_N}\, d^N\l\,\,
 \overline{\Phi}_{\l}(x)\,\,\, \Phi_{\l}(x'+n).}
The last equality follows from the following property
of the  eigenfunctions \NORMWFUNCAFF\ of the $N$-particle Hamiltonian in
Nonlinear Schr\"{o}dinger theory:
\eqn\NORMWFUNCAFFF{
\Phi_{\l}(x+n)=
\sum_{w\in W}(-1)^{l(w)}\,
\prod_{i<j}\left(\frac{\l_{w(i)}-\l_{w(j)}+ic}
{\l_{w(i)}-\l_{w(j)}-ic}\right)^{n_i}
\exp(2\pi i\sum_m\l_{w(m)}n_m)\times
}
$$
\times \prod_{i<j}\left(\frac{\l_{w(i)}-\l_{w(j)}+ic}
{\l_{w(i)}-\l_{w(j)}-ic}\right)^{\frac{1}{2}+[x_i-x_j]}
\exp(2\pi i\sum_k\l_{w(k)}x_k),$$
These wave functions are periodic and descend to the wave
functions \NORMWFUNC\
 if $\l=(\lambda_1,\cdots ,\lambda_N)$ satisfy \BAONE.

The representation for the kernel \PROJKERNELPERIODD\ leads to the
following
representation for the Green function (cylinder path integral) and
torus partition function for $U(N)$ Yang-Mills-Higgs theory for $t_{i\neq
2}=0$:
\eqn\cylo{G_{YMH}(x,x')=}
$$=\sum_{n \in \IZ^N} \int_{\IR^N/S_N}d^N\l\,\,\mu(\l)\,\,
\overline{\Phi_{\lambda}(x)}
\,e^{2\pi  i \sum_{m=1}^N\l_m n_m} \,\,
\prod_{l\neq j}
\left(\frac{\l_l-\l_j-ic}{\l_l-\l_j+ic}\right)^{n_j}e^{-t_2p_{2}(\l)}\Phi_{\lambda}(x'),
$$
or same:
$$G_{YMH}(x,x')=\sum_{k\in \IZ^N}G^0_{YMH}(x,x'+k)=\sum_{k\in \IZ^N}
\int_{\IR^N/S_N}\, d^N\l\,\,
 \overline{\Phi}_{\l}(x)e^{-t_2p_2(\l)} \Phi_{\l}(x'+k).$$
Similarly for kernel: $\widetilde K(x,x') =\sum_{k\in \IZ^N}
\widetilde K_0(x,x'+k)$  since the kernel is a Green function at $t_2=0$.
 Integrating over $x$ after setting $x=x'$
we obtain the representation for the partition function on the torus:
\eqn\tornew{Z_{YMH}(\Sigma_1)=}
$$=
\int_{\IR^N/S_N}d^N\l\,\,\mu(\l)\,\,\sum_{(n_1,\cdots ,n_N)\in \IZ^N}
e^{2\pi  i \sum_{m=1}^N\l_m n_m} \,\,
\prod_{l\neq j}
\left(\frac{\l_l-\l_j-ic}{\l_l-\l_j+ic}\right)^{n_j}e^{-t_2p_{2}(\l)}.$$
This is in a complete agreement with
a representation for the partition function
 of $U(N)$ Yang-Mills-Higgs theory  on a torus discussed in Section 3,
formula \SDHPART.
   All above expressions have the property to
recover corresponding well-known answers of 2d YM theory
in the limit $c \rightarrow \infty$,
as they should from the general arguments presented before.
Note that one can repeat the same arguments for
all observables and higher differential operators of Nonlinear
Schr\"{o}dinger theory,  traces of
higher powers of Dunkle operator from Section 4,
by simply turning on all other couplings $t_{k}$.

The identification of the representation of the partition function
of Nonlinear Schr\"{o}dinger operator and Yang-Mills-Higgs theory on
the torus  strongly suggests that the full equivalence of the
theories.  It would be very desirable to obtain same answer for Green
function and  wave-function (cylinder path integral)
directly from path integral for cylinder topology using  the
cohomology localization technique.

Finally let us comment on  the explicit form of the wave function
\NORMWFUNCAFFF\ form the gauge theory point of view.
The appearance of an integer part $[x]$ in  \NORMWFUNCAFFF\
is not quite unexpected phenomena form the point of view
of the proposed identification of the wave functions
in Yang-Mills-Higgs theory with the wave functions in Nonlinear
Schr\"{o}dinger theory.  Let  us remark  that the interpretation of the
topologically
non-trivial bundles as an interpolation between topologically nontrivial
gauge transformations naturally arises in the discussion of the
spectral flow of  the eigenfunctions of the
gauge invariant operators on the  boundary (see
\ref\SUW{E.~Witten, {\it Global gravitational anomalies},
Commun. Math. Phys. {\bf 100} (1985) 197-226.},
 \ref\AETA{M.~Atiyah, {\it The logarithm of the Dedekind
     $\eta$-function}, Math. Ann. {\bf 278} (1987) 335-380.}
for the details and examples).
To make a closer contact with this interpretation let us recall
the simplest instance of
Atiyah-Patodi-Singer index theorem on an two-dimensional manifold
with non-empty  boundary \ref\APS{M. F. ~Atiyah, V. K. ~Patodi,
  I. M. ~Singer {\it Spectral Asymmetry and Riemannian Geometry},
 Math. Proc. Camb. Philos. Soc. {\bf 77}
  (1975) 43, {\bf 78} (1975) 403, {\bf 79} (1976) 71.}.

Let us given  an even-dimensional
Riemann  spin manifold $M$ with a  boundary $\pr M$, and a vector bundle
$\CE$ supplied with a  connection $\nabla _A$. Consider  Dirac
operator on a vector bundle $\CE\otimes S$ where  $S$ is a spinor bundle
supplied with a connection $\nabla_{S}$.
 The definition of the index of the Dirac
operator on a non-compact manifold $M$ relies on the correct
treatment of the boundary conditions.
In  \APS\  specific non-local boundary conditions were defined
corresponding to a vacuum state  in the Hilbert space
of the Dirac fermions. We would like to apply
the index theorem to the two-dimensional cylinder
$L$,  $\pr L=S^1_+\cup S^1_-$  with a flat metric
and $U(1)$ bundles supplied with a connection $\nabla_A$.
The spinor bundle on $L$  can be  identified with $\Omega^{0,0}(L)\oplus
\Omega^{0,1}(L)$ and the Dirac operator is given by
$\CD=\apr_A+\apr_A^{+}$.
We have the following expression for the index of $\CD$
which coincides with the index of $\apr_A$:
\eqn\INDEX{
{\rm Index}_{\CD}=\int_L c_1(\nabla)+\frac{1}{2}\eta( S^1_+\cup S^1_-)}
where the $\eta$-invariant $\eta(S^1_+\cup S^1_-)$ is
defined in terms of the spectrum of the
restriction of $\CD$ to the boundary :
\eqn\EATINV{
\eta(\pr L)=\lim_{s\to 0}\,\sum_{\l_i \in {\rm Spec}(\CD|_{\pr L})}
{\rm sign}(\l_i)\,
|\l_i|^s.}
Taking into account the relation  $\int_L F(A)=\int_{S_+^1}A
-\int_{S_-^1}A$ we have:
\eqn\DIFERETA{
(\eta(S^1_+)-\frac{i}{2\pi}\int_{S^1_+}A)-(\eta(S^1_-)-\frac{i}{2\pi}\int_{S^1_-}A)
={\rm Index}_{\CD}
\in \IZ.}
An easy calculation shows that the $\eta$-invariant for a
 a constant  connection $\nabla_A=\pr_t+x$, on $S^1$ is given by:
\eqn\ETASONE{
-\frac{1}{2}\eta(S^1)=x-[x]-\frac{1}{2}. }
where $[x]$ is an integer part of $x$. Thus:
 \eqn\ETACON{
\frac{i}{2\pi}\int_{S^1}A-\frac{1}{2}\eta(S^1)=-x+(x-[x]-\frac{1}{2})=[x]+\frac{1}{2}.}
This seems makes the appearance of integer values in \NORMWFUNCAFFF\
less mysterious. One should stress however   that the proper derivation
of the wave function \NORMWFUNCAFFF\ using this reasoning
 should use a more refined form of the
$\eta$-invariant also introduced in \APS.
 Let $B$ be  an operator commuting with
$\CD$. Then character-valued $\eta$-invariant is given by:
\eqn\ETACHAR{
\eta_B(\pr M)=
\lim_{s\to 0}\,\sum_{\l_i \in {\rm Spec}(\CD|_{\pr M})}
 {\rm sign}(\l_i)\,|\l_i|^s\,{\rm tr}_{V_{\l_i}}\, B,}
where $V_{\l_i}$ is an eigenspace of $\CD$ corresponding to the eigenvalue
$\l_i$.
Using an appropriate operator $B$ one can reproduce the phase factor in
\NORMWFUNCAFFF.

Finally note that  $\eta$-invariant  is defined using the
vacuum boundary condition on the quantum fields  \APS.
Thus the appearance of $\eta$-invariant in the explicit expression for a
wave function  can be traced back to the
 the fact that the additional fields
$(\Phi,\psi_{\Phi},\varphi_{\pm},\chi_{\pm})$
 entering  the description of
the Yang-Mills-Higgs theory are in the vacuum state. Thus the
only contribute  to the total wave function is a  phase factor.

\newsec{On equivariant cohomology description of the Hilbert space}

The realization of the representations of the degenerate affine Hecke
algebras $\CH_{\Fg,c}$ in the space of the $S^1\times G$-equivariant
cohomology of the flag spaces  for the Lie group $G$, such that
$\Lie(G)=\Fg$
 considered in \ref\Lu{G.~Lusztig, {\it Equivariant $K$-theory
 and Representations of Hecke Algebras},
Proc. Am. Math. Soc. {\bf 94} (1985) 337.},
\ref\KL{D.~Kazhdan, G.~Lusztig, {\it Proof
of the Deligne-Langlands conjecture for Hecke algebras},
Invent. Math. {\bf 87} (1987) 153.} (see also
\ref\GinC{V.~Ginzburg, N.~Chriss, {\it Representation theory and
    complex geometry}, Birkhauser, Boston, 1997.},
\ref\GinB{V.~Ginzburg, {\it Geometric methods in representation
theory of Hecke algebras and quantum groups},
 Notes by Vladimir Baranovsky. NATO Adv. Sci. Inst. Ser. C
 Math. Phys. Sci., 514, 14, [arXiv:math/9802004].}),
bears an obvious resemblance with the constructions discussed in
this paper.  Below we make some preliminary remarks regarding this
relation. The detailed consideration will be postponed for another
occasion.

We start with the simplest case of the dimensionally reduced
Yang-Mills theory. The one-dimensional Yang-Mills theory in the Hamiltonin
formulation is a system with a first class constraint and thus the Hilbert
space is naturally described as a cohomology of the corresponding BRST
operator. The standard
approach is to realize a Hilbert space as the cohomology of the BRST
operator acting in some extended space including ghost variables.
One considers  a pair of canonically conjugate
ghost-antighost fields $(b,c)$ of ghost numbers $(-1,1)$
such that BRST operator acts on the
space of functions of $\varphi$ and $c$, and is given by:
\eqn\BRSTTOP
{Q_{BRST}=\Tr\,(c\,[\varphi,\frac{\pr}{\pr
    \varphi}])-\frac{1}{2}\Tr\,([c,c],\frac{\pr}{\pr c}).}
The Hilbert space $\CH=H^*_{Q_{BRST}}$  of the theory is naturally
graded by the ghost number.  The relevant
cohomology can be interpreted as a Lie algebra cohomology of
$\Fg=\Lie(G)$ with coefficients in the  space of functions on $\Fg$
considered as a $\Fg$-module with respect to the adjoint action:
\eqn\HILCOH{
 \CH= H^*_{Q_{BRST}}=H^*(\Fg,{\rm Fun}(\Fg)).}
Note that the higher cohomologies are
non-trivial and result differs from the naive expectation of finding the
space of  the gauge invariant function on $\Fg$ as a realization of
the Hilbert space.  Let us remark that \HILCOH\ is close to $H_G(G)^*$.

One can try to use a more economical way to quantize the
theory. Let us consider the same set of fields but use a  modified BRST
operator:
\eqn\BRSTTOPMOD
{Q_{BRST}=Tr\,(c\,[\varphi,\frac{\pr}{\pr
    \varphi}])-\frac{1}{2}Tr\,([c,c],\frac{\pr}{\pr
    c})+\varphi\frac{\pr}{\pr c}.}
The cohomology of this BRST operator provides a BRST model
for $G$-equivariant cohomology of the point (see \ref\KAL{J.~Kalkman,
{\it BRST Model for Equivariant Cohomology and Representative
for Equivariant Thom Class}, Commun. Math. Phys. {\bf 153} (1993)
477.} for discussion of various models for equivariant cohomology):
\eqn\NEWHS{\CH=H^*_G(pt)\equiv H^*(BG)=\IC[G]^G}
We see that in this formulation we get the correct Hilbert space.

The interpretation of the Hilbert space of the dimensionally reduced
Yang-Mills theory
in terms of the equivariant
cohomology is very natural. Let us recall that according
to \ref\FHT{D.~Freed, M.~Hopkins, C.~Teleman, {\it
 Twisted K-theory and loop group representations},
[arXiv:math/0312155].} the Hilbert space of the
$G/G$ gauged Wess-Zumino-Witten  theory at the level $k$
can be described in terms of  twisted
equivariant  cohomology $K_G(G)^{*+k+c_v+\dim(G)}$. In the limit $k\to
\infty$ this provides a description of the states in Yang-Mills  theory
in terms of equivariant cohomologies $H^*_G(G)$. After the reduction
to one dimensions one gets the equivariant cohomology
$H^*_G(G)\otimes \IC$ with coefficients in $\IC$ which provide a model
for Hilbert space of Yang-Mills  theory.

\newsec{ On Nahm transform and Langlands duality}

Taking into account previous considerations it is natural to look
for more direct correspondence between two-dimensional $U(N)$
Yang-Mills-Higgs theory and quantum non-relativistic Nonlinear
Schr\"{o}dinger theory associated with $U(2)$ (
Nonlinear Schr\"{o}dinger theory exists for any group see e. g. \FT,
here we only utilized the $U(2)$ version of it).
In this section we briefly comment on this issue leaving
the detailed considerations for the future work.

Let us  consider a  covariant description of the phase space of
the theory as a space of  solutions of the equations of motions
 (see  \ref\CWCOVH{C.~Crnkovic,
 E.~Witten {\it Covariant description of canonical formalism
in geometric theories} Newton's tercentenary volume,
edit. S.~Hawking  and W.~Israel, Cambridge University Press 1987.}
for discussion of the general formalism).  Applying
the general construction to the two-dimensional theory one can
identify the phase space  with a  space of classical solutions on
$S^1\times \IR$. Consider first the Yang-Mills theory.
Equations of motion for $\varphi$ lead to the flatness condition
$F(A)=0$ and thus the moduli space of the classical solutions has a
natural projection onto  the space of unitary flat $G$-connections on
 $S^1\times \IR$. Under  appropriate boundary conditions
this space can  be identified with $G/{\rm Ad}_G=H/W$.
The fiber of the projection is given by the space of covariantly constant
sections $\nabla_A\, \varphi=0$ and thus the total phase space can be
identified with the cotangent bundle to a moduli space of flat
connections. Thus, indeed the  covariant phase space coincides
with the  phase space $\CM=T^*H/W$ described in Section 2.

Consider now Yang-Mills-Higgs theory for $c=0$.  Equations of motion
obtained by the variation over $(\varphi_0,\varphi_+,\varphi_-)$  are
given by:
\eqn\COVPH{F(A)-\Phi\wedge
  \Phi=0,\,\,\,\,\,\,\nabla_A^{(1,0)}\Phi^{(0,1)}=0,\,\,\,\,\,\,
\nabla_A^{(0,1)}\Phi^{(1,0)}=0,}
and those given by the variation over $(A,\Phi)$  are:
$$
\nabla_A^{(1,0)}\varphi_+=c\Phi^{(1,0)}+[\Phi^{(1,0)},\varphi_0],$$
 \eqn\EQMOTADD{
\nabla_A^{(0,1)}\varphi_-=-c\Phi^{(0,1)}+[\Phi^{(0,1)},\varphi_0].}
 $$
\nabla_A\varphi_0=[\Phi^{(1,0)},\varphi_+]-[\Phi^{(0,1)},\varphi_-].$$
Let us start with   the  space $\CM_H$  of the solutions of the first set
of equations.
The  equations:
\eqn\PARTCOMB{
F(A)-\Phi\wedge \Phi=0,\,\,\,\,\,\,\,\,\,\nabla_A\Phi=0,}
are equivalent to a flatness condition of the modified
connection:
\eqn\FALTCOND{
(d+A+i\Phi)^2=0,}
and have a simple solution on $\IC^*=S^1\times\IR$:
\eqn\FLATCONDA{
A^c=A+i\Phi=g_c^{-1}dg_c+g_c^{-1}A_D^cg_c,}
where $A^c_D$ is constant one form taking values in the diagonal matrices.
It is useful to represent complex matrix $g_c$ as $g_c=bg$ where $b\in
GL(N,\IC)/U(N)$
is a Hermitian  matrix and $g\in U(N)$ (Cartan decomposition).
Then $g$ can be gauged away and we have:
\eqn\FLATCONDB{
A=A^b_D-(A^b_D)^+,\,\,\,\,\,\,\Phi=-i(A^b_D+(A^b_D)^+),}
where $A^b_D=b^{-1}db+b^{-1}A^c_Db$.
The third equation from the first set provides a constraint
on $b$ which fixes it up to a holomorphic map $\IC^*=S^1\times
\IR\rightarrow G$
(harmonicity condition). Variant of Narasimhan-Seshadri and
 Ramanathan arguments \ref\NSONE{M. S.~Narashimhan, C. S.~Seshadri, {\it
Stable and unitary vector bundles on a compact Riemann surface},
 Ann. of Math. (2) {\bf 82} (1965), 540.},
\ref\RAM{M.S.~Ramanathan, {\it Stable bundles on a compact Riemann
surface}, Math. Ann. {\bf 213} (1975) 129.}
  allow to describe,  in the holomorphic
terms, the moduli space of flat $G$-bundles for a compact complex curve.
Consider the equation $\nabla_A^{(0,1)}\Phi^{1,0}=0$. It describes
a holomorphic section of the holomorphic bundle. The rest of the equations
define  the unitary structure and have unique solutions on the compact
surface. In the non-compact case  the  same arguments
work for appropriate boundary conditions.
Thus, we can think of the phase space of  Yang-Mills-Higgs theory
as a moduli space of  Hitchin equations on $\IC^*=S^1\times \IR$.

Now we can try to apply  Nahm duality to characterize
this  moduli space in other terms. Indeed, the
 solutions of Hitchin equations  on $S^1\times \IR$ can be considered as
solutions of four-dimensional  (anti)self-dual  YM equations on
$\IR\times
S^1_{R_1}\times
S^1_{R_2}\times S^1_{R_2}$  when $R_2\to 0$.
By Nahm duality \ref\NAHMD{W. ~Nahm, {\it
The construction of all self-dual multimonopoles
by the ADHM method}
, in "Monopoles in quantum field theory", eds. N. Craigie,
World Scientific, Singapore, 1982.}, \ref\NAHMDD{
 W.~Nahm, {\it Self-dual monopoles and calorons},
 Lect. Notes in Physics. 201, eds. G.
Denardo,  (1984) p. 189.}
the moduli space of (anti)self-dual gauge fields  on
$S^1_{R_0}\times S^1_{R_1}\times S^1_{R_2}
\times S^1_{R_2}$  for a gauge group $U(N)$  is equivalent to the moduli
space of (anti)self-dual gauge fields
on $S^1_{1/R_0}\times S^1_{1/R_1}\times S^1_{1/R_2}\times S^1_{1/R_2}$
for another gauge group $U(M)$ where $M$ is second Chern
class of the gauge field (instanton number).
  Therefore taking $R_0\to
\infty$, $R_2\to 0$ we get  an equivalence of the moduli
space of solutions of Hitchin equations for $G=U(N)$ with the
 moduli space of periodic monopoles  on $S_{R_1}^1\times \IR\times
 \IR$ with the gauge group $U(M)$ where $M$ is
an appropriately defined  topological characteristic
of  the solutions with fixed boundary conditions.  Taking into account
that
the cylinder has two boundaries the simplest nontrivial boundary
conditions leads to $M=2$.  Thus, one can expect
that the moduli space for $U(N)$-gauge theory
will be equivalent to $N$-monopole solution in a $U(2)$ gauge theory.
These considerations correspond to the case $c=0$. One can hope that
considering $S^1$-equivariant version of the Nahm  correspondence one
obtains the analogous  relations  for $c\neq 0$.  This would provide
a hint for  the  direct connection between
Yang-Mills-Higgs theory for $G=U(N)$  and  Nonlinear Schr\"{o}dinger
theory associated with $U(2)$.

Let us finally note  that under Nahm  duality  $G$-bundles
on a  four dimensional torus
$S^1_{R_0}\times S^1_{R_1}\times S^1_{R_2}\times S^1_{R_3}$
map to  $G^{L}$-bundles on the dual torus
$S^1_{1/R_0}\times S^1_{1/R_1}\times S^1_{1/R_2}\times S^1_{1/R_3}$
where $G^L$ is a  Langlands dual group (at least for classical groups).
 Thus one would expect that the proposed relation
between Yang-Mills-Higgs theory and Nonlinear Schr\"{o}dinegr theory
being generalized to the case of an arbitrary semisimple Lie group
should be related to the  Langlands duality.

\newsec{Generalization  of $G/G$ gauged WZW model}

It is natural to expect
that the story presented in previous sections extends from
the Yang-Mills-Higgs  theory
to the certain generalization  of $G/G$ gauged WZW theory. In this section
we describe such construction and show that the
 partition function can be represented as a sum
over solutions of a certain generalization of Bethe Ansatz equation.

We start with the definition of the
set of fields and the action of the odd and even symmetries
in the spirit of \UONEACT, \GAUGEYM, \BRST. Let us note that
the gauged Wess-Zumino-Witten  model can be obtained
from the topological Yang-Mills theory by using the group-valued field $g$
instead of algebra-valued field $\varphi$. Correspondingly, we
replace the generators of the Lie algebra  actions with the
parameter $\varphi$ \BRSTYM, \GAUGEYM\
by the generators of the Lie group action \GAUGEYMGWZW, \BRSTYMGWZW\
with the parameter $g$. Thus it is natural to introduce the
set of fields $(A,\psi_A,\Phi,\psi_{\Phi},\chi_{\pm},\varphi_{\pm},g)$
and $t\in \IR^*$ with the following action of the
odd and even symmetries:
\eqn\EVENGGWZW{
\CL_{(g,t)}\,A^{(1,0)}=(A^g)^{(1,0)}-A^{(1,0)}
\,\,\,\,\,\,\,\CL_{(g,t)}\,A^{(0,1)}_A=-(A^{g^{-1}})^{(0,1)}+A^{(0,1)},}
$$\CL_{(g,t)}\,\psi_A^{(1,0)}=-g\psi_A^{(1,0)}
g^{-1}+\psi_A^{(1,0)},\,\,\,\,\,\,
\CL_{(g,t)}\,\psi_A^{(0,1)}=g^{-1}\psi_A^{(0,1)}
g-\psi_A^{(0,1)},\,\,\,\,\CL_{(g,t)}\,g=0,$$
$$\CL_{(g,t)}\,\Phi^{(1,0)}=tg\Phi^{(1,0)}
g^{-1}-\Phi^{(1,0)},\,\,\,\,\,\,
\CL_{(g,t)}\,\Phi^{(0,1)}=-t^{-1}g^{-1}\Phi^{(0,1)} g+\Phi^{(0,1)},
$$
$$
\CL_{(g,t)}\,\psi_{\Phi}^{(1,0)}=tg\psi_{\Phi}^{(1,0)}
g^{-1}-\psi_{\Phi}^{(1,0)},
\,\,\,\,\,\,
\CL_{(g,t)}\,\psi_{\Phi}^{(0,1)}=-t^{-1}g^{-1}\psi_{\Phi}^{(0,1)}
g+\psi_{\Phi}^{(0,1)},
$$
$$\CL_{(g,t)}\,\chi_{+}=tg\chi_+g^{-1}-\chi_+,\,\,\,\,\,\,
\CL_{(g,t)}\chi_-=-t^{-1}g^{-1}\chi_-g+\chi_- $$
$$
\CL_{(g,t)}\,\varphi_{+}=t^{-1}g\varphi_{+}g^{-1}-\varphi_{+},\,\,\,\,\,\,
\CL_{(g,t)}\varphi_{+}=-tg^{-1}\varphi_{+}g+\varphi_{+}
$$
\eqn\ODDGGWZW{
Q\,A=i\psi_A,\,\,\,\,\,\,Q\,\psi^{(1,0)}_A=i(A^g)^{(1,0)}-iA^{(1,0)},
\,\,\,\,\,\,\,\,Q\,\psi^{(0,1)}_A=-i(A^{g^{-1}})^{(0,1)}+iA^{(0,1)},}
$$Q\,g=0,$$
\eqn\BRSTGGWZW{
Q\Phi=i\psi_{\Phi},\,\,\,\,
Q\psi^{(1,0)}_{\Phi}=t g\Phi^{(1,0)} g^{-1}-\Phi^{(1,0)},
\,\,\,\,\,Q\psi^{(0,1)}_{\Phi}=-t^{-1}g^{-1}\Phi^{(0,1)} g-\Phi^{(0,1)},}
$$Q\chi_{\pm}=i\varphi_{\pm},\,\,\,\,
Q\varphi_+=t g\chi_+ g^{-1}- \chi_+,\,\,\,\,\,\,
Q\varphi_-=-t^{-1} g^{-1}\chi_- g+ \chi_- .$$
We have $Q^2=\CL_{(g,t)}$ and $Q$ can be considered as a
BRST operator on the space of $\CL_{(g,t)}$-invariant
functionals.

We define the action of the theory in analogy with the construction
of the action for Yang-Mills-Higgs theory as follows:
\eqn\ACTONQCOM{
S=S_{GWZW}+[Q,\int_{\Sigma_h}\,d^2z \,\Tr\,( \frac{1}{2}
\Phi\wedge  \psi_{\Phi}+}
$$+\tau_1\,(\varphi_+
 \nabla^{(1,0)}_A\Phi^{(0,1)}+\varphi_-
 \nabla^{(0,1)}_A\Phi^{(1,0)})+\tau_2(\chi_+\varphi_-+\chi_-\varphi_+)
 {\rm vol}_{\Sigma_h})]_+.$$
Taking $\tau_1=0$, $\tau_2=1$ and
 applying  the  standard localization technique to this theory
 we obtain for the partition function:
\eqn\GWZWHPART{
Z_{GWZWH}(\Sigma_h)=\frac{e^{(1-h)a(t)}}{|W|}
\int_{H} d^N\l\,\,\mu_q(\l)^{h}\,\sum_{(n_1,\cdots,n_N)\in \IZ^N}
e^{2\pi i \sum_{m=1}^N\l_m n_m(k+c_v)}\times }
$$\times  \prod_{j\neq k}(e^{2\pi i(\l_j-\l_k)}-1)^{n_j-n_k+1-h}
\,\,\prod_{j,k}(te^{2\pi i(\l_j-\l_k)}-1)^{n_j-n_k+1-h},$$
where $a(t)$ is a $h$-independent constant, the integral
goes over the Cartan  torus $H=(S^1)^N$ and
\eqn\QMEASURE{
\mu_q(\l)=\det\|\frac{\pr \beta_j(\l)}{\pr \l_k}\|
,}
with:
\eqn\newone{e^{2\pi i \beta_j(\lambda)}=e^{2\pi i
\l_j(k+c_v)}\,\prod_{k\neq
  j}\frac{te^{2\pi i (\l_j-\l_k)}-1}{te^{2\pi i (\l_k-\l_j)}-1}.}
We can rewrite this formula in the form similar to
\SDHsumon:
\eqn\GWZWHPARTo{
Z_{GWZWH}(\Sigma_h)=\frac{e^{(1-h)a(t)}}{|W|}
\int_{H} d^N\l\,\,\mu_q(\l)^{h}\,\sum_{(n_1,\cdots,n_N)\in \IZ^N}
e^{2\pi i \sum_{m=1}^N \beta_m(\lambda) n_m}\times }
$$\times  \prod_{j< k}(e^{i\pi(\l_j-\l_k)}-e^{i\pi(\l_k-\l_j)})^{2-2h}
\,\,\prod_{j<k}|te^{i\pi (\l_j-\l_k)}-e^{i\pi(\l_k-\l_j)}|^{2-2h}.$$
Summation over
integers in \GWZWHPART\ leads to the following restriction on the
integration parameters:
\eqn\QBA{
e^{2\pi i \l_j(k+c_v)}\,\prod_{k\neq
  j}\frac{te^{2\pi i (\l_j-\l_k)}-1}{te^{2\pi i (\l_k-\l_j)}-1}=1
\,\,\,\,,\,\,\,\,\,\,\, i=1,\cdots ,N,}
It is useful to rewrite the equations \QBA\ in the standard form
of the Bethe Ansatz equations:
\eqn\QBAsin{
e^{2\pi i \l_j(k+c_v)}\,\prod_{k\neq
  j}\frac{sin(i\pi (\l_j-\l_k +ic))}{sin(i\pi (\l_j-\l_k-ic))}=1
\,\,\,\,,\,\,\,\,\,\,\, i=1,\cdots ,N,}
This clearly shows that we are dealing with a kind of XXZ quantum
integrable chain. The particular form \QBAsin\ can be obtained
by the taking the limit $s\to -i\infty$ in the following Bethe
equations:
\eqn\QBAsin{
\left(\frac{sin(i\pi (\l_j-i s c ))}{sin(i\pi (\l_j+i s c))}\right)
^{(k+c_v)}\,\prod_{k\neq
  j}\frac{sin(i\pi (\l_j-\l_k +ic))}{sin(i\pi (\l_j-\l_k-ic))}=1
\,\,\,\,,\,\,\,\,\,\,\, i=1,\cdots ,N,}
corresponding to formal limit of the infinite spin $s$ of XXZ chain.

The partition function is the generalization of Yang-Mills-Higgs
theory, discussed above, and can be written in the following form:
\eqn\PARTDISCRQ{
Z_{GWZWH}(\Sigma_h)=\sum_{\l_i\in \CR_q}\,\,(D_{\l}^q)^{2-2h},}
where $\CR_q$ is a set of the solutions of \QBA\ and:
\eqn\SDHDIMQ{
D^q_{\l}=\mu_q(\l)^{-1/2}
\prod_{i<j}(q^{\frac{1}{2}(\l_i-\l_j)}-q^{\frac{1}{2}(\l_j-\l_i)})
\prod_{i<j}|t q^{\frac{1}{2}(\l_i-\l_j)}-q^{\frac{1}{2}(\l_j-\l_i)}|,}
where  we use the standard parametrization $q=\exp(2\pi i/(k+c_v))$.
Note that in the limit $t\to \infty$ equation \QBA\
and the expression for the partition function \SDHDIMQ\
up to an  overall scaling factor become the corresponding expressions
for a  gauged Wess-Zumino-Witten model. Finally note that the form of
\QBA\
and the explicit expressions for the q-Casimir operators, playing the
role of the Hamiltonians, strongly imply the description of the wave
functions of the theory in terms of the wave functions in a particular
$XXZ$ finite spin chain. This proposition will be discussed in details
elsewhere.

\newsec{Conclusion}

Let us put  the results of this paper in a more general perspective.
Any two-dimensional topological theory satisfying the appropriate
cutting/gluing relations \ref\Atiyah{M. F.~Atiyah, {\it Topological
quantum field theory},  Publications Mathmatiques de l'IH\'{E}S,
{\bf 68} (1988), p. 175-186.}  can be described by a
commutative Frobenius algebra. Generally  this Frobenius algebra
comes from the chiral ring $\CR$  of some Conformal Field Theory
(CFT). For example the gauged WZW model leads  to
finite-dimensional Frobenius algebra associated with the representation
theory of the finite-dimensional quantum groups. The corresponding
CFT is a WZW model. Similarly, the two-dimensional
topological Yang-Mills theory is related to  the
infinite-dimensional Frobenius algebra constructed  in terms of the
representation theory of finite-dimensional Lie groups. The
associated CFT is a particular degeneration of the
WZW model. Thus, one should expect that  the
topological Yang-Mills-Higgs theory introduced in \mns\ and its
generalizations defined in this paper correspond to some interesting
classes of  two-dimensional CFT. One can speculate that such CFT
should be constructed by an appropriate deformation of the
WZW model for complex gauge groups. Taking into
account the relation between WZW models for complex
groups  and the (generalizations) two-dimensional quantum gravity
this might be elaborated in precise form.

In this paper we mostly restrict ourselves to the case $G=U(N)$. This
restriction is not essential. One can study these theories for
an arbitrary semisimple Lie group
(for the  construction of  finite-particle wave functions for
arbitrary $G$ see e. g. \Ga).

Let us stress that the simple expression for the partition function
as a sum over the solutions of Bethe Ansatz equations can be
considered as a kind of nonlinear Fourier transform of the similar,
but more conventional, representation as a sum over a set of
$\IC^*$-fixed point components of the Higgs bundle moduli spaces
(see \mns\ for details). The equality of the  two dual
representations of the partition function can be considered as an
example of the Arthur-Selberg trace formula if one takes into
account the interpretation of $\IC^*$-fixed points as variations of
the Hodge structures on the underlying curve. Note also that the
existence of the two dual representations leads to  sum rules for
the solutions of the Bethe Ansatz equations and the corresponding
nonlinear Fourier transform looks very similar to the Quantum
Inverse Scattering Method for quantum integrable systems \SFT. It
would be interesting to make this analogy more precise. Non-linear
Fourier transforms become standard tools in the study of topological
sigma models and mirror symmetry and here we see another important
appearance.  Let us note that it would be also interesting to
establish the relation between results of the present paper and the
other type of connection between 2d Yang-Mills theory and many-body
systems described in \ref\GN{A. Gorsky, N. Nekrasov, {\it
Hamiltonian systems of Calogero type and two dimensional Yang-Mills
theory}, Nuclear Physics B, {\bf 414}, [arXiv:hep-th/9304047]}.

Given a $U(N)$ gauge theory it is natural to consider its t'Hooft
limit $N\to \infty$. In this limit one expects to find a dual
description of the theory in terms of  stringy expansion. The known
results for the dual string description of two-dimensional
Yang-Mills theory \ref\gt{D. J.~
  Gross, W.~Taylor, {\it Two Dimensional QCD is a String Theory,}
Nucl.Phys. {\bf B400} (1993) 181-210.}, \ref\stef{S.~Cordes, G.~Moore,
S.~Ramgoolam, {\it Large N 2D Yang-Mills Theory and Topological
String Theory,} Commun.Math.Phys. 185 (1997) 543-619,
[arXiv:hep-th/9402107].}, \ref\stefone{S.~Cordes, G.~Moore,
S.~Ramgoolam, {\it Lectures on 2D Yang-Mills Theory, Equivariant
Cohomology and Topological Field Theories, } Nucl.Phys.Proc.Suppl.
41 (1995) 184-244, [arXiv:hep-th/9411210].} implies that the similar
description of $N\to\infty$ limit of Yang-Mills-Higgs theory and its
generalizations can be very instructive. In this respect let us note
that recently the same program has been completed for $q$-deformed
two-dimensional Yang-Mills theory  \ref\Vafa{M.~Aganagic, H.~Ooguri,
N.~Saulina, C.~Vafa,  {\it
Black Holes, q-Deformed 2d Yang-Mills, Non-perturbative
Topological Strings}, [arXiv:hep-th/0411280].}.
 The $q$-deformed theories are closely related
to the gauged WZW deformations of two-dimensional
 Yang-Mills theory. Thus the deformation of the Yang-Mills-Higgs
 theory proposed in Section 10 can provide a dual description of
 interesting string backgrounds.
In this respect it is interesting to note that the proper
quasi-classical expansion of the Bethe Ansatz solutions, given via
the double scaling limit $c \rightarrow 0, N \rightarrow \infty$
with $Nc$-fixed, is known to lead to Riemann surfaces  and genus
expansion \ref\fedya{F.~Smirnov, {\it Quasi-classical Study of Form
Factors in Finite Volume,} [arXiv:hep-th/9802132].}.

The  gauge theories considered in this paper  seem to
 provide  a proper framework for the quantum field theory version of
a Kazhdan-Lusztig type construction of the representations of
double affine Hecke algebras (DAHA).  Indeed the essential role
played by $S^1\times G$-equivariant cohomology groups and $K$-groups
 both in the Kazhdan-Lusztig  constructions and in the quantum gauge
theory construction implies a deep relation between these two
subjects. In the topological gauge theories considered here the
natural objects are gauge invariant and thus reproduce only the
properties of the center of DAHA. One can expect that
the  consideration of the corresponding  CFT
discussed above should provide a more deeper connection with the
representation theory of DAHA. This seems compatible with the ideas
regarding the relation between the representation theory of affine Hecke
algebra and hyperk\"{a}hler geometry of Higgs bundles advanced in
\ref\AB{M.~Atiyah,   R.~Bielawski, {\it Nahm's equations, configuration
spaces and flag
manifolds}, [arXiv:math/0110112].}.

Note that the Hecke algebras are important ingredients of the
explicit construction of the Langlands correspondence (see
\ref\Bump{D.~Bump {\it Automorphic Forms and Representations},
  Cambridge  Univ. Press, Cambridge, 1998.},\ref\LLL{
{\it An Introduction to the Langlands Program}, ed. J.~Bernstein,
S.~Gelbart, Birkhauser,  2003.} for modern introductions into the
subject). Thus it is tempting to suggest that the relation between
gauge theories and double affine Hecke algebras considered in this
paper can lead to a better understanding of this correspondence.
The latter might be related to recent studies in
\ref\KW{A.~Kapustin, E.~Witten,
 {\it Electric-Magnetic Duality And The Geometric Langlands Program},
  [arXiv:hep-th/0604151].}  where an interpretation of the geometric
Langlands correspondence in terms of quantum field
theory was described.

{\bf  Acknowledgements}.  We are grateful to J. Bernstein, M.
Kontsevich, W. Nahm, N. Nekrasov, F. Smirnov and L. Takhtajan for
discussions.
The research of the first author  is partly supported by the
grant RFBR 04-01-00646 and Enterprise Ireland Basic Research Grant;
 that of the second author is supported by Enterprise Ireland Basic
Research Grant, SFI Research Frontiers Programme and Marie Curie RTN
ForcesUniverse from EU.

\appendix{A}{Twistor type description}

In this Appendix we give a Chern-Simons type  representation
of the bosonic part of the Yang-Mills-Higgs action using a twistor
approach.
Consider the one-dimensional  family of flat connections:
\eqn\TWone{
\CD=d+\CA(\l)=d+\frac{1}{\l}\Phi^{(1,0)}+A+\l\,\Phi^{(0,1)}.}
The
flatness condition for $\CD$:
 \eqn\TWtwo{ \CF(\CA)\equiv \CD^2=0.}
is equivalent to the Hitchin equations for the pair $(A,\Phi)$. Then we
have the following representation for the bosonic part of the
action $S_0$ \ACTBOS:
\eqn\TWfour{ S(A,\Phi)=Res_{\l=0}\int\,d^2z
\,Tr(\varphi(z,\zb,\l)\CF(\CA(z,\zb,\l)) +c
\,\CA(z,\zb,\l)\,\frac{\pr}{\pr \l}\, \CA(z,\zb,\l)).} Here
$\varphi(z,\zb,\l)=\varphi_+(z,\zb)+\l^{-1}\varphi_0(z,\zb)+\l^{-2}\varphi_-(z,\zb)$.
Let us note  that  the elements of the  gauge group
 are considered to be independent of $\l$.
 In this representation it is natural to denote
$\varphi(z,\zb,\l)\equiv \CA_{\l}(z,\zb,\l)$ to get,
formally, a three-dimensional action. Note however that
there are severe constraints on the dependence of the fields on
$\l$. Thus, it might be considered as a reduction of three dimensional
Chern-Simons theory.

\appendix{B}{ On Lagrangian geometry  of the singular manifold}

In this Appendix we discuss  the classical gauge theory counterpart of
the $c$-dependent boundary conditions \BCN\ used for the construction of
the eigenfunctions in  Nonlinear Schr\"{o}dinger theory. We will
consider only the case of $G=U(2)$ in the  gauge theory
reduced to one dimension. The phase space in this case
is $\CM=T^*\IR^2/\IZ_2$. Choose
 the coordinates $(x_1,x_2,\pi_1,\pi_2)$  on $T^*\IR^2$ such that
$(x_1,x_2)$
are coordinates on $\IR^2$ and $(\pi_1,\pi_2)$ are coordinates on
the fibers of the projection $T^*\IR^2\to \IR^2$. Let  $X=x_2-x_1$,
 $Y=\pi_2-\pi_1$ and the action of $\IZ_2$  be given by:
\eqn\Bone{
  w_0:(X,Y)\rightarrow (-X,-Y).}
The space of functions on the phase space $\CM$
is generated  by the invariant functions  $a_1=\frac{1}{2}X^2$,
$a_2=\frac{1}{2}Y^2$  and $a_3=XY$
satisfying:  \eqn\Btwo{
F(a)=4a_1a_2-a_3^2=0.} One has $a_1\geq 0$, $a_2\geq 0$ and the fiber of
the projection on the subspace $a_1>0$, $a_2>0$ consists of two
points while the fiber over $(a_1=0) \cup (a_2=0)$ consists of one
point. The point $a_1=a_2=0$ is singular (i.e. $F=0$, $\pr_iF=0$ at
this point). The quantization of this space in the standard
representation $\hat{X}=x$, $\hat{Y}=\pr_x$ is given by the
representation of the algebra ${\frak sl}(2)$:
\eqn\Bthree{
[\hat{a}_2,\hat{a}_1]=\hat{a}_3,\,\,\,\,[\hat{a}_3,\hat{a}_1]
=2\hat{a}_1,\,\,\,\,\,[\hat{a}_3,\hat{a}_2]=-2\hat{a}_2.}
where:
\eqn\QUNATSLTWO{
\hat{a}_1=\frac{1}{2}x^2,\,\,\,\,\,\hat{a}_2=\frac{1}{2}\pr_x^2,\,\,\,\,\
\hat{a}_3=\frac{1}{2}(x\pr_x+\pr_x x).}
The
equation $4a_1a_2-a_3^2=0$ translates into the restriction on the
  value of the  second Casimir operator  in this representation.
 Thus we are dealing here  with the representations  associated
with the nilpotent orbits of ${\frak sl}(2)$.
The representations can be
described in terms of the  space of  regular functions on Lagrangian
submanifold $\CL_0$ defined by the equation  $a_2=0$. The functions
 on $\CL_0$ are the functions of $a_1$ ($a_3=0$ on $\CL_0$) and
the action of $\hat{a}_i$ is given by:
\eqn\BBone{ \hat{a}_1f(a_1)=a_1f(a_1),}
\eqn\BBtwo{
\hat{a}_2f(a_1)=\frac{\pr f(a_1)}{\pr a_1}+4a_1\frac{\pr^2 f}{\pr
a_1^2},}
\eqn\BBthree{ \hat{a}_3f(a_1)=4a_1\frac{\pr f(a_1)}{\pr
a_1}+\frac{1}{2}f(a_1).} Regular functions depending on $x^2$  can be
characterized by
the following conditions: \eqn\BBfour{
\delta(x)\pr_x(x^{-1}\pr_x)^nf(x)=0, \,\,\,\,\,\,\, n\geq 0.} It
easy to see that these conditions are invariant with respect to the
action of $\IZ_2$ and compatible with the action of ${\frak
sl}(2)$. The first condition can be written in a more usual form as:
$$\pr_Xf(X)|_{X=0}=0$$
which is the boundary condition discussed above.

Now consider the case $c\neq 0$.
The change of the boundary condition can be understood in the geometric
terms as follows.  As it was discussed previously the phase
space is defined by the equations:
\eqn\singular{
a_1a_2-a_3^2=0,\,\,\,\,\,a_1\geq 0,\,\,\,a_2\geq 0.}
  In the case of the Yang-Mills theory
we quantize the cone \singular\ using the Lagrangian submanifold
$\CL_0$ defined by the equation $a_2=0$. This Lagrangian submanifold
is rather special. Note that the associated
 representation of ${\frak sl}(2)$ is irreducible. Its naive deformation
$\CL_{c}$
defined by the equation $a_2=c^2$ provides {\it reducible}
representation. To understand it better consider the lift
$\widetilde{\CL}_{c}$ of $\CL_c$ to  $(x,y)$-plane. The lift
 of $\CL_0$ is given by the connected submanifold $x=0$. On the other
hand the lift $\widetilde{\CL}_{c\neq 0}$ of $\CL_{c\neq 0}$
consists of two components $y=\pm c$. This means that the
representation associated with $\CL_{c\neq 0}$ can be reducible.
Indeed one can choose as a $\IZ_2$-invariant Lagrangian submanifold
the submanifold  $\hat{\CL}^+_{c\neq 0}$  defined  by the equations:
\eqn\ONETWOTHReEE{
 \widetilde{\CL}_{c\neq 0}^+=(y=+c,x>0)\cup (y=-c,x<0).}
Taking the  factor over $\IZ_2$ we obtain:
 \eqn\BBBone{ \CL_{c\neq
0}^+=(a_1>0,a_3>0,a_2=c^2).}
Let us describe the corresponding
spaces of functions and the action of the generators of ${\frak
sl}(2)$. We  start with the space of functions on
$\widetilde{\CL}_{c}$ where $c\neq 0$. As functions of $x$ it is
given by a pair of functions with the action of $\hat{y}$ by the
following differential operator:
\eqn\BBBtwo{ \hat{y}\pmatrix{f_+(x)
\cr f_-(x)}=\pmatrix{\pr_x+c & 0 \cr 0 & \pr_x-c} \pmatrix{f_+(x)
\cr f_-(x)},} and the action of the generator $w_0$ of $\IZ_2$ is
given by: \eqn\BBBthree{ w_0 \pmatrix{f_+(x) \cr
f_-(x)}=\pmatrix{f_-(-x) \cr f_+(-x)}.}
Equivalently we have:
 \eqn\BBBfour{
\hat{y}\pmatrix{f_+(x)e^{cx} \cr f_-(x)e^{-cx}}=\pmatrix{\pr_x & 0
\cr 0 & \pr_x}\pmatrix {f_+(x)e^{cx} \cr f_-(x)e^{-cx}},} and
\eqn\BBBfive{ w_0 \pmatrix{f_+(x)e^{cx} \cr
f_-(x)e^{-cx}}=\pmatrix{f_-(-x)e^{cx} \cr f_+(-x)e^{-cx}}.}
Thus the symmetry condition is reduced to  $f_+(x)=f_-(-x)$.
Note that we automatically have $f_+(0)=f_-(0)$. Now consider the
functions defined on the lift of the Lagrangian submanifold $\CL^+_c$.
They can be described as follows: \eqn\BBBBone{
f(x)=f_+(x)e^{cx},\,\,\,\,\,\,\,\,\,\,\,\,\,x>0,} \eqn\BBBBtwo{
f(x)=f_+(-x)e^{-cx},\,\,\,\,\,x<0.}
The symmetry condition implies that $f_+(x)$ is a symmetric function. Note
that:
$f(x)|_{x\to +0}=f(x)|_{x\to -0}$ and $\pr_xf(x)|_{x\to +0}=cf(0)$,
$\pr_xf(x)_{x\to -0}=-cf(0)$. Thus we have the {\it deformed }
boundary conditions. As an example consider the function that is a
solution of the equation $\pr_x^2f(x)=\l^2f(x)$ for $x\neq 0$ and
satisfy the boundary conditions at $x=0$:
\eqn\BBBBA{
f_{\l}(x)=(\l-ic)\,e^{i\l x} +(\l+ic)\,e^{-i\l x},\,\,\,\,\, x>0,}
\eqn\BBBBAA{ f_{\l}(x)=(\l+ic)\,e^{i\l x} +(\l-ic)\,e^{-i\l
x},\,\,\,\,\, x<0.}
 This function  can be easily represented in the
form discussed above:
\eqn\BBBBAAA{
f_{\l}(x)=\left((\l-ic)\,e^{i(\l+ic) x} +(\l+ic)\,e^{-i(\l-ic)
    x}\right)e^{cx},
\,\,\,\,\, x>0,} \eqn\BBBBAAAA{ f_{\l}(x)=
\left((\l+ic)\,e^{i(\l-ic) x} +(\l-ic)\,e^{-i(\l+ic)
x}\right)e^{-cx},\,\,\,\,\, x<0.}
\listrefs
\end